\documentclass[sigconf,screen,nonacm]{acmart}

\author{Yisu Remy Wang}
\affiliation{
  \institution{University of Washington}
  \country{USA}
}
\email{remywang@cs.washington.edu}

\author{Max Willsey}
\affiliation{
  \institution{University of Washington}
  \country{USA}
}
\email{mwillsey@cs.washington.edu}

\author{Dan Suciu}
\affiliation{
  \institution{University of Washington}
  \country{USA}
}
\email{suciu@cs.washington.edu}

\keywords{Worst-case optimal join}

\usepackage{tikz}
\usepackage[inline]{enumitem}
\usepackage[font=small,labelfont=bf]{caption}
\usepackage{subcaption}
\usepackage{algorithm}
\usepackage[noend]{algpseudocode}

\usepackage{bm} 

\usepackage{graphics}
\graphicspath{{_build/}{fig/}}

\usepackage[colorinlistoftodos]{todonotes}

\usepackage{listings} 

\usepackage{xspace}

\newcommand{\FJ}{\textsf{Free Join}\xspace}
\newcommand{\GJ}{\textsf{Generic Join}\xspace}
\newcommand{\WCOJ}{\textsf{WCOJ}\xspace}
\newcommand{\GHT}{\textsf{GHT}\xspace}
\newcommand{\GHTs}{\textsf{GHT}s\xspace}
\newcommand{\COLT}{\textsf{COLT}\xspace}
\newcommand{\COLTs}{\textsf{COLT}s\xspace}
\newcommand{\cd}{\text{ :- }}
\newcommand{\setof}[2]{\{{#1}\mid{#2}\}}        
\newcommand{\set}[1]{\{#1\}}        

\newcommand{\imdbavgfjbj}{2.94x\xspace}
\newcommand{\imdbavgfjgj}{9.61x\xspace}
\newcommand{\imdbmaxfjbj}{19.36x\xspace}
\newcommand{\imdbmaxfjgj}{31.6x\xspace}
\newcommand{\imdbminfjbj}{0.85x\xspace}
\newcommand{\imdbmaxbjfj}{17\%\xspace}
\newcommand{\imdbminfjgj}{2.63x\xspace}

\begin{document}

\title{\FJ: Unifying Worst-Case Optimal and Traditional Joins}

\begin{abstract}
  Over the last decade, worst-case optimal join (\WCOJ) algorithms have
emerged as a new paradigm for one of the most fundamental challenges
in query processing: computing joins efficiently.  Such an algorithm
can be asymptotically faster than traditional binary joins, all the
while remaining simple to understand and implement.  However, they
have been found to be less efficient than the old paradigm,
traditional binary join plans, on the typical acyclic queries found in
practice.  Some database systems that support \WCOJ use a hypbrid
approach: use \WCOJ to process the cyclic subparts of the query (if
any), and rely on traditional binary joins otherwise.  In this paper
we propose a new framework, called \FJ, that unifies the two
paradigms.  We describe a new type of plan, a new data structure
(which unifies the hash tables and tries used by the two paradigms),
and a suite of optimization techniques.  Our system, implemented in
Rust, matches or outperforms both traditional binary joins and \GJ on
standard query benchmarks.
\end{abstract}

\ccsdesc[500]{Information systems~Join algorithms}

\maketitle

\section{Introduction}\label{sec:intro}

Over the last decade, worst-case optimal join (\WCOJ)
algorithms~\cite{DBLP:conf/pods/NgoPRR12, DBLP:conf/icdt/Veldhuizen14,
  DBLP:journals/sigmod/NgoRR13, DBLP:conf/pods/000118} have emerged as
a breakthrough in one of the most fundamental challenges in query
processing: computing joins efficiently.  Such an algorithm can be
asymptotically faster than traditional binary joins, all the while
remaining simple to understand and
implement~\cite{DBLP:journals/sigmod/NgoRR13}.  These algorithms have
opened up a flourishing field of research, leading to both theoretical
results~\cite{DBLP:journals/sigmod/NgoRR13,DBLP:conf/pods/Khamis0S17}
and practical
implementations~\cite{DBLP:conf/icdt/Veldhuizen14,DBLP:journals/tods/AbergerLTNOR17,DBLP:journals/pvldb/FreitagBSKN20, DBLP:journals/pvldb/MhedhbiS19}.

Over time, a common belief took hold: 
  ``\WCOJ is designed for cyclic queries''.
This belief is rooted in the observation that 
  \WCOJ enjoys lower asymptotic complexity 
  than traditional algorithms for cyclic queries~\cite{DBLP:journals/sigmod/NgoRR13}, 
  but when the query is acyclic, 
  classic algorithms like the Yannakakis algorithm~\cite{DBLP:conf/vldb/Yannakakis81} 
  are already asymptotically optimal. 
Moreover, traditional binary join algorithms have benefited from 
  decades of research and engineering.
Techniques like column-oriented layout, vectorization, 
  and query optimization
  have contributed compounding constant-factor speedups,
  making it challenging for \WCOJ to be competitive in practice.
This has lead many instantiations of \WCOJ,
  including Umbra~\cite{DBLP:journals/pvldb/FreitagBSKN20},
   Emptyheaded~\cite{DBLP:journals/tods/AbergerLTNOR17}, and Graphflow~\cite{DBLP:journals/pvldb/MhedhbiS19},
  to adopt a hybrid approach: 
  using \WCOJ to process parts of the query,
  and resorting to traditional algorithms (usually binary join) 
  for the rest.
Having two different algorithms in the same system
  requires changing and potentially duplicating existing infrastructure 
  like the query optimizer. 
This introduces complexity, and hinders the adoption of \WCOJ.

The dichotomy of \WCOJ versus binary join has led researchers 
  and practitioners to view the algorithms as opposites.
In this paper, we break down this dichotomy by proposing
  a new framework called \FJ that unifies \WCOJ and binary join.
%
We propose several new techniques to make \FJ outperform
both binary join and \WCOJ:
we design an algorithm to convert any binary join plan to a \FJ plan
that runs as fast or faster; we design a new data structure called
\COLT (for \emph{Column-Oriented Lazy Trie}), adapting the classic
column-oriented layout to improve the trie data structure used in
\WCOJ; and we propose a vectorized execution algorithm for \FJ.

\begin{figure}
    \centering
    \includegraphics[width=0.75\linewidth]{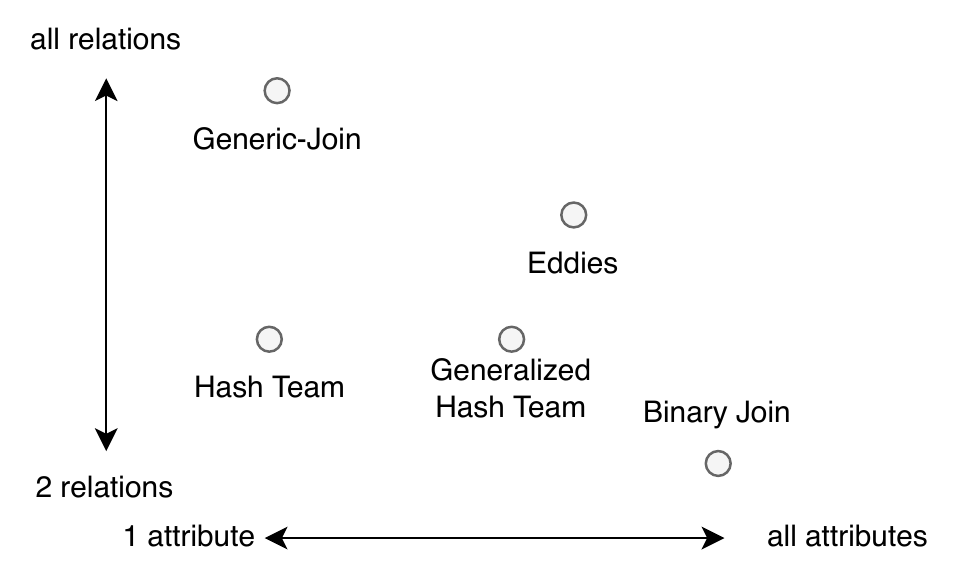}
    \caption{Design space of join algorithms.}
    \label{fig:design-space}
\end{figure}

To explain these contributions we provide some context.
In this paper, we focus on algorithms based on hashing,
  and choose \GJ~\cite{DBLP:journals/sigmod/NgoRR13} as a representative of \WCOJ algorithms.
A crucial difference between \GJ and binary join lies 
  in the way they process each join operation. 
Binary join processes two relations at a time, 
  and joins on \emph{all attributes} 
  in the join condition between these two relations. 
In contrast, \GJ processes one attribute at a time, 
  and joins \emph{all relations} that share that attribute.
This suggests a design space of join algorithms, 
  where each join operation may process any 
  number of attributes and relations.
Figure~\ref{fig:design-space} shows this design space
  which also covers classic multiway join algorithms 
  like Hash Team~\cite{DBLP:conf/vldb/GraefeBC98}, 
  Generalized Hash Team~\cite{DBLP:conf/vldb/KemperKW99}
  and Eddies~\cite{DBLP:conf/sigmod/HellersteinA00}.
Being able to join on any number of variables and relations
  frees us from the constraints of all existing algorithms 
  mentioned above. 

Our new framework, \FJ,
  covers the entire design space, 
  thereby generalizing and unifying existing algorithms.
The starting observation is that the execution of a left-deep linear 
  binary join plan is already very similar to \GJ.
While \GJ (reviewed in Sec.~\ref{sec:background}) is traditionally specified as a series of nested loops~\cite{DBLP:journals/sigmod/NgoRR13},
  the push-based model~\cite{DBLP:journals/pvldb/Neumann11,DBLP:journals/pvldb/KerstenLKNPB18} for executing a left-deep linear binary plan
  is also implemented, similarly, as nested loops.
The two algorithms also process each join operation similarly:
  each binary hash join iterates over tuples on one relation, 
  and for each tuple probes into the hash table of another relation;
  each loop level in \GJ iterates over the keys of a certain trie,
  and probes into several other tries for each key.
This inspired us to unify hash tables and hash tries into the same data structure, 
  and develop \FJ using iteration and probing as key operations.
This finer-grained view of join algorithms allows \FJ
  to generalize and unify existing algorithms,
  while precisely capturing each of them.

  \FJ takes as input an already optimized binary join plan, and
  converts it into a new kind of plan that we call an \FJ plan.  It
  then optimizes the \FJ plan, resulting in a plan that sits in
  between binary join and \GJ, combining the benefits of both.  On one
  hand \FJ takes full advantage of the design space in
  Figure~\ref{fig:design-space}.  On the other hand, by starting from
  an already optimized binary plan, \FJ takes advantage of existing
  cost-based optimizers; in our system we used binary plans produced by
  the optimizer of DuckDB~\cite{DBLP:conf/cidr/RaasveldtM20,DBLP:conf/vldb/Raasveldt22}.

Next, we address the main source of
  inefficiency in \GJ: the need to construct a trie on each relation
  in the query.  In contrast,  
a binary join plan needs to build a hash map only for each relation 
  on the right-hand side of a join,
  and simply iterates over the relation on the left.
In practice, trie-building has been observed to be a major 
  bottleneck for \GJ~\cite{DBLP:journals/pvldb/MhedhbiS19,DBLP:journals/pvldb/FreitagBSKN20}, 
  making it slower than binary join.
This is because each trie is more expensive to build than a hash map,
  and the left relation is usually chosen to be a large relation 
  by the query optimizer.
One simple optimization in \FJ is that we do not built a trie for
tables that are left children, mimicking the binary plans.
However, we go a step further, and introduce the {\em Column-Oriented
  Lazy Trie} (\COLT) data structure, which builds the inner subtries
  lazily, by creating each subtrie on demand.  
We note that this builds on an earlier idea
in~\cite{DBLP:journals/pvldb/FreitagBSKN20}.  As the name suggests,
\COLT adapts the lazy trie data structure
in~\cite{DBLP:journals/pvldb/FreitagBSKN20} to use a column-oriented
layout.  And unlike the original lazy trie which builds at least one
trie level per table, \COLT completely eliminates the cost of trie
building for left tables.

Finally, we describe a method for incorporating vectorized
  processing in \FJ, allowing it to collect multiple data values
  before entering the next iteration level.  
The standard \GJ processes one data value at a time, but, as is the
  case in traditional query engines, this leads to poor cache
  locality.  
Vectorized execution~\cite{DBLP:conf/icde/PadmanabhanAMJ01} was proposed for binary join
  to improve its locality by processing data values in batch.
By breaking down join operations into iterations and probes, 
  \FJ gives rise to a simple vectorized execution algorithm
  that breaks each iteration into chunks and groups together 
  batches of probes.
Our proposal is to our knowledge the first vectorized execution algorithm for \GJ.

To evaluate \FJ, we implemented it as a standalone Rust library, and
compared it with two baselines:
  \begin{enumerate*}
    \item our own \GJ implementation in Rust, and
    \item the binary hash join implemented in
      DuckDB~\cite{DBLP:conf/cidr/RaasveldtM20,DBLP:conf/vldb/Raasveldt22},
      a state-of-the-art in-memory database.
    \end{enumerate*}
  We found that, on acyclic queries, \FJ is up to \imdbmaxfjbj
  faster than binary join, and up to \imdbmaxfjgj faster than \GJ; on
  cyclic queries, \FJ is up to 15.45x faster than binary join, and up
  to 4.08x faster than \GJ.

  While optimizers for binary plans have been developed and improved
  over decades~\cite{DBLP:conf/sigmod/SelingerACLP79}, little
  is known about optimizing \GJ.  A \GJ plan consists of a total order
  on its variables, and its run time does depend on the choice of this
  order.  But since the theoretical analysis of \GJ guarantees worst
  case optimality for {\em any} variable order, it is a folklore
  belief that \GJ is more robust than binary join plans to poor
  choices of the optimizer.  We also conduced experiments measuring
  the robustness of the three types of plans (binary, \GJ, \FJ) to
  poor choices of the optimizer.  We found that \GJ is indeed the
  least sensitive, while \FJ, like binary joins, suffers more from the
  poor optimization choices of the optimizer, since both rely on a
  cost-based optimized plan.  However, \GJ starts from worse baseline
  than \FJ.  In other words, \FJ takes better advantage of a good
  plan, when available, than \GJ does.

In summary, we make the following contributions in this paper:
\begin{enumerate}
  \item \FJ, a framework unifying existing join algorithms (Section~\ref{sec:free-join}).
  \item An algorithm to converting any binary join plan into an
    optimized \FJ plan (Section~\ref{sec:bj-to-fj}).
  \item \COLT, a column-oriented lazy trie data structure (Section~\ref{sec:colt}).
  \item A vectorized execution algorithm for \FJ (Section~\ref{sec:vectorized-execution}).
  \item Experiments evaluating the algorithms and optimizations (Section~\ref{sec:eval}).
\end{enumerate}
\section{Background}\label{sec:background}

This section defines basic concepts and reviews background on binary
join and \GJ.

\subsection{Basic Concepts}\label{sec:basic-concepts}

We consider a relational database where each relation has a fixed
\emph{schema}, and may have duplicates, i.e. we use bag semantics.
A \emph{full conjunctive query} has the following form:
\begin{align}
  Q(\bm x) \cd & R_1(\bm x_1), \ldots, R_m(\bm x_m). \label{eq:cq}
\end{align}
Each term $R_i(\bm x_i)$ is called an \emph{atom}, where $R_i$ is a
relation name and $\bm x_i$ a tuple of variables.  The query is
\emph{full}, meaning that the head variables $\bm x$ include all
variables appearing in the atoms.  To reduce clutter in the following
sections, we will assume that the query does not have self-joins.
This is without loss of generality: if two atoms have the same
relation name, then we simply rename one of them.  Our system also
supports selections, projections, and aggregation.  We assume that the
selections are pushed down to the base tables, thus the atom $R_i$
in~\eqref{eq:cq} may include a selection over a base table; in
particular, all variables in the atom $R_i(\bm x_i)$ are distinct.
Similarly, projections and aggregates are performed after the full
join, hence none of them is shown in~\eqref{eq:cq}.

\begin{example} \label{ex:triangle} Consider the following SQL query:
\begin{lstlisting}[
    language=SQL,
    showspaces=false,
    basicstyle=\ttfamily\small,
    % numbers=left,
    % numberstyle=\tiny,
    commentstyle=\color{gray}
 ]
SELECT r.x, s.u, t.u
  FROM R as r, M as s, M as t -- schema: R(x,y), M(u,v,w)
 WHERE s.w > 30 AND t.v = t.w
   AND r.y = s.u AND s.v = t.u AND t.v = r.x
\end{lstlisting}
Then we denote by $S = \Pi_{uv}(\sigma_{w>30}(M))$ and $T =
\Pi_{uv}(\sigma_{v=w}(M))$, and write the query as:
$$Q_{\triangle}(x,y,z) \cd R(x, y), S(y,z), T(z, x).$$
We call this query the \emph{triangle query} over the relations $R$, $S$,
$T$.
\end{example}

It is often convenient to view the conjunctive query~\eqref{eq:cq} as
a hypergraph.  The \emph{query hypergraph} of $Q$ consists of vertices
$\mathcal{V}$ and edges $\mathcal{E}$, where the set of nodes
$\mathcal{V}$ is the set of variables occurring in $Q$, and the set of
hyperedges $\mathcal{E}$ is the set of atoms in $Q$.  The hyperedge
associated to the atom $R(\bm x_i)$ is defined as the set consisting
of the nodes associated to the variables $\bm x_i$.  As standard, we
say that the query $Q$ is {\em acyclic} if its associated hypergraph
is $\alpha$-acyclic~\cite{DBLP:journals/jacm/Fagin83}

\subsection{Binary Join}\label{sec:binary-join}

The standard approach to computing a conjunctive query~\eqref{eq:cq} is
to compute one binary join at a time.  A {\em binary plan} is a binary
tree, where each internal node is a join operator $\Join$, and each
leaf node is one of the base tables (atoms) $R_i(\bm x_i)$ in the
query~\eqref{eq:cq}.  The plan is a \emph{left-deep linear plan}, or
simply left-deep plan, if the right child of every join is a leaf
node.  If the plan is not left-deep, then we call it \emph{bushy}.
For example, $(R \Join S) \Join (T \Join U)$ is a bushy plan, while
$((R \Join S) \Join T) \Join U$ is a left-deep plan.  We do not treat
specially right-deep or zig-zag plans, but simply consider them to be
bushy.

In this paper we consider only hash-joins, which are the most
common types of joins in database systems. 
The standard way to execute a bushy plan is to
decompose it into a series of left-deep linear plans.  Every join node
that is a right child becomes the root of a new subplan, which is
first evaluated, and its result materialized, before the parent join
can proceed.  As a consequence, every binary plan, bushy or not,
becomes a collection of left-deep plans. We decompose bushy
plans in exactly the same way, and we will focus on left-deep linear
plans in the rest of this paper.  For example, the bushy plan
$(R \Join S) \Join (T \Join U)$ is converted into two plans:
$P_1 = T \Join U$ and $P_2 = (R \Join S) \Join P_1$; both are
left-deep plans.

To reduce clutter, we represent a left-deep plan
$(\cdots ((R_1 \Join R_2) \Join R_3) \cdots \Join R_{m-1}) \Join R_m$
as $[R_1, R_2, \ldots, R_m]$.  Evaluation of a left-deep plan is done
using pipelining.  The engine iterates over each tuple in the
left-most base table $R_1$; each tuple is probed in $R_2$; each of the
matching tuple is further probed in $R_3$, etc.

\begin{figure}
  \begin{subfigure}[b]{0.45\linewidth}
\begin{lstlisting}
for (x, y) in R:
  s = S[y]?
  for (y, z) in s:
    t = T[x,z]?
    for (x, z) in t:
      output(x, y, z)
\end{lstlisting}
    \caption{Binary join.}
    \label{fig:background:binary-join}
  \end{subfigure}
  \begin{subfigure}[b]{0.45\linewidth}
    \centering
\begin{lstlisting}
for a in R.x $\cap$ T.x:
  r = R[a]; t = T[a]
  for b in r.y $\cap$ S.y:
    s = S[b]
    for c in s.z $\cap$ t.z:
      output(a, b, c)
\end{lstlisting}
    \caption{\GJ.}
    \label{fig:background:gj}
  \end{subfigure}
  \caption{Execution of binary join and \GJ for $Q_\triangle$.  The
    notation \lstinline|S[y]?| performs a lookup on $S$ with the key
    $y$, and continues to the enclosing loop if the lookup fails.
    Binary join iterates over {\em tuples}, \GJ iterates over {\em
      values}.}
\end{figure}

\begin{example}
  A possible left-deep linear plan for $Q_\triangle$ is $[R, S, T]$,
  which represents $(R(x,y) \Join S(y,z)) \Join T(z,x)$.  To execute
  this plan, we first build a hash table for $S$ keyed on $y$, where
  each $y$ maps to a vector of $(y,z)$ tuples, 
  and a hash table for $T$ keyed on $x$ and $z$, each mapped to a
  vector of $(x,z)$ tuples\footnote{When the relations are bags, then
    the hash table may contain duplicate tuples, or store separately
    the multiplicity.  We also note that the question what exactly to
    store in the hash table (e.g. copies of the tuples, or pointers to
    the tuple in the buffer pool) has been studied for a long time,
    see~\cite{DBLP:journals/csur/Graefe93}.}.  Then the execution
  proceeds as shown in Figure~\ref{fig:background:binary-join}.  For each
  tuple $(x, y)$ in $R$, we first probe into the hash table for $S$
  using $y$ to get a vector of $(y, z)$ tuples.  We then loop over
  each $(y, z)$ and probe into the hash table for $T$ using $x$ and
  $z$.  Each successful probe will return a vector of $(x, z)$ tuples,
  and we output the tuple $(x, y, z)$ for each $(x, z)$.
\end{example}

\subsection{\GJ}\label{sec:background:gj}

\GJ was introduced in~\cite{DBLP:journals/sigmod/NgoRR13} and is the
simplest worst-case optimal join algorithm.  It is based on the
earlier Leapfrog Triejoin algorithm~\cite{DBLP:conf/icdt/Veldhuizen14}.
\GJ computes the query $Q$ in~\eqref{eq:cq} through a series of nested
loops, where each loop iterates over a variable (not a tuple).
Concretely, \GJ chooses arbitrarily a variable $x$, computes the
intersection of all $x$-columns of all relations containing $x$, and
for each value $a$ in this intersection it computes the residual query
$Q[a/x]$, where every relation $R$ that contains $x$ is replaced with
$\sigma_{x=a}(R)$.  In pseudocode:
\begin{lstlisting}
GJ: for a in $\bigcap \setof{\Pi_x(R_i)}{R_i \mbox{ contains } x}$
      compute Q[a/x]  \\ run GJ on Q with one fewer variable
\end{lstlisting}
If the query $Q$ has $k$ variables, then there are $k$ nested loops in
\GJ.  In the inner most loop, \GJ outputs the tuple of constants, one
from each iteration.\footnote{For bag semantics, it multiplies their
  multiplicities.}  We notice that a plan for \GJ consists of a total
order of the variables of the query, which we denote as
$[x_1, x_2, \ldots, x_k]$.  Assuming that the intersection above is
done optimally (see below), the algorithm is provably
worst-case-optimal, for any choice of the variable order.

\begin{example}
  Fig.~\ref{fig:background:gj} illustrates the pseudocode for \GJ on the
  query $Q_\triangle$, using the variable order $[x,y,z]$.  We denoted
  $\Pi_x(R)$ by $R.x$, and denoted (with some abuse) $\sigma_{x=a}(R)$
  by $R[a]$.
\end{example}

While binary joins use hash tables, an implementation of \GJ uses a
\emph{hash trie}, one for each relation in the query.  The hash-trie
is a tree, whose depth is equal to one plus the number of attributes
of the relation, and where each node is either an empty leaf
node,\footnote{For bag semantics, we store in the leaf the
  multiplicity of the tuple.} or a hash map mapping each atomic value
to another node.  We will call the \emph{level} of a node to be the
distance from the root, i.e. the root has level 0, its children level
1, etc.  The hash-trie completely represents the relation: every
root-to-leaf path corresponds precisely to one tuple in the relation.
\GJ uses the hash-trie as follows.  In order to compute
$\sigma_{x=a}(R)$, it simply probes the current hash table for the
value $x=a$, and returns the corresponding child.  To compute an
intersection $\Pi_x(R_1) \cap \Pi_x(R_2) \cap \cdots$, it selects the
trie with the fewest keys, say $R_1$, then iterates over every value
$a$ in the keys for $R_1$ and probes it in each of the hash-maps for
$R_2, R_3, \ldots$; this is a provably optimal algorithm for the
intersection.

\begin{example}
  Consider the query $Q_\triangle$ and the \GJ plan $[x, y, z]$.  We
  first build a hash trie each for $R$, $S$, and $T$.  Each trie has
  three levels including the leaf.  Level 0 of $R$ is keyed on $x$,
  level 1 is keyed on $y$, level 2 contains empty leaf nodes, and
  similarly for $S$ and $T$.  Consider again the pseudocode in
  Figure~\ref{fig:background:gj}.  The first loop intersects level 0
  of the $R$-trie and the $T$-trie.  For each value $a$ in the
  intersection, we retrieve the corresponding children $R[a]$ and
  $T[a]$ respectively; these are at level 1.  The second loop
  intersects the hash map $R[a]$ (at level 1) with the level 0
  hash-map of $S$.  For each value $b$ in the intersection it
  retrieves the corresponding children (levels 2 and 1 respectively),
  and, finally, the innermost loop intersects the $S$- and $T$-hash
  maps (both at level 2), and outputs $(a,b,c)$ for each $c$ in the
  intersection.  So far we have assumed set semantics; if the
  relations have bag semantics, then we simply multiply the tuple
  multiplicities on the leaves (level 3).
\end{example}


\subsection{Binary Join v.s. \GJ}
Binary join and \GJ each have their own advantages and disadvantages.
\GJ became popular because of its asymptotic performance guarantee:
\citet{DBLP:journals/sigmod/NgoRR13} proved the algorithm is
\emph{worst-case optimal} for \emph{any variable order}, in the sense
that its run time is bounded by the largest possible size of its
output, called AGM bound~\cite{DBLP:journals/siamcomp/AtseriasGM13}.
For example, \GJ executes $Q_\triangle$ in time
$\sqrt{|R|\cdot |S| \cdot |T|}$, which is $n^{3/2}$ when all relations
have size $n$; in contrast, a binary join plan can take $\Omega(n^2)$.
We note, however, that this formula does not include the preprocessing
time needed to construct the tries.  For example, if $T$ is
significantly larger than $R, S$, then the run time of \GJ is
$\ll |T|$, yet during preprocessing \GJ needs to read the entire
relation $T$.  On the other hand, binary join has been a staple of
database systems for decades.  The hash table data structure is
simpler than hash tries and is cheaper to build.  Techniques like
vectorized execution and column-oriented layout have also made binary
join practically efficient, but these optimizations have not been
adapted for \GJ.  Binary join plans are known to be very sensitive to
the choice of the optimizer: poor plans perform catastrophically
bad~\cite{DBLP:journals/pvldb/LeisGMBK015}.  In contrast, although the
runtime performance of \GJ does depend on the variable order, some
researchers believe that \GJ is less sensitive to poor variable
orders, in part because it is always theoretically optimal.

\section{\FJ}\label{sec:free-join}

In this section we introduce the \FJ framework.
We start by presenting the Generalized Hash Trie (\GHT) which
 is the data structure used in \FJ (Section~\ref{sec:ght}).
Next we introduce the \FJ plan that specifies 
  how to execute a query with \FJ (Section~\ref{sec:fj-plan}).
Finally, we describe the \FJ algorithm, 
  which takes as input a collection of \GHTs 
  and a \FJ plan,
  and computes the query according to the plan (Section~\ref{sec:free-join-algorithm}).
  
We will show how each of the above components generalizes 
  and unifies the corresponding components in binary join 
  and \GJ: 
  the \GHT generalizes hash tables and hash tries,
  the \FJ plan generalizes binary plans and \GJ plans, and
  the \FJ algorithm generalizes binary join and \GJ.

\begin{figure}
\begin{align*}
  Q_{\clubsuit}(x,a,b,c) :- R(x,a),S(x,b),T(x,c)
\end{align*}
\begin{align*}
  R & = \set{(x_0, a_0)} \cup \setof{(x_1, a_i^l), (x_2, a_i^r)}{i \in [1 \ldots n]} \\
  S & = \set{(x_0, b_0)} \cup \setof{(x_2, b_i^l), (x_3, b_i^r)}{i \in [1 \ldots n]} \\
  T & = \set{(x_0, c_0)} \cup \setof{(x_3, c_i^l), (x_1, c_i^r)}{i \in [1 \ldots n]} 
\end{align*}
\caption{The clover query $Q_\clubsuit$, and an input instance.}
\label{fig:clover-query}
\end{figure}

\begin{figure}
  \centering
  \includegraphics[width=0.7\linewidth]{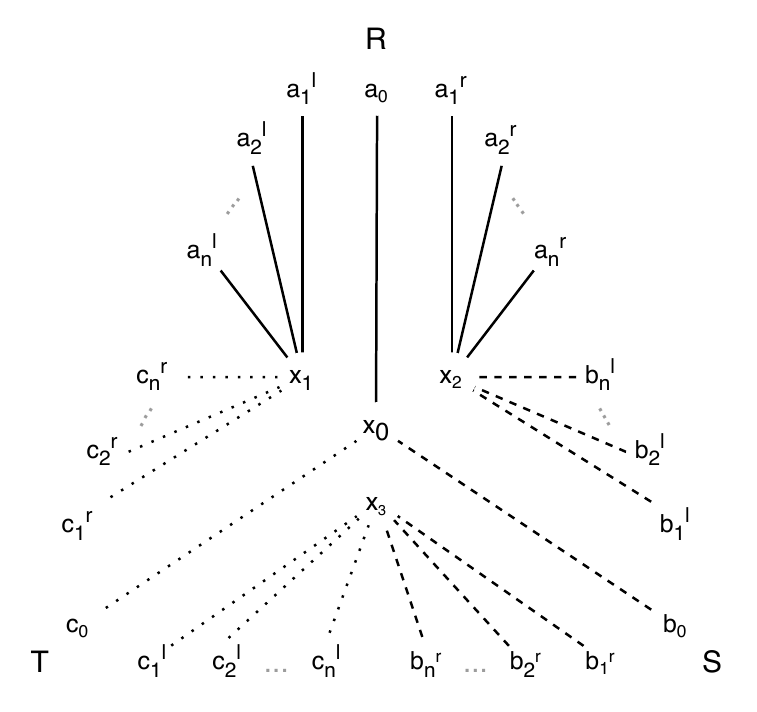}
  \caption{Visualization of the instance in
    Fig.~\ref{fig:clover-query}. The solid (top) edges form the relation
    \texttt{R}, the dashed (right) edges form the relation \texttt{S},
    and the dotted (left) edges form the relation \texttt{T}.  The
    relations join on the attribute in the center.  The only output
    tuple consists of the three edges in the center.  }
  \label{fig:clover-vis}
\end{figure}

Throughout this section we will make use of the {\em clover query}
$Q_\clubsuit$ in Figure~\ref{fig:clover-query}.
Figure~\ref{fig:clover-vis} visualizes the input relations for this query.
Note that $Q_\clubsuit$ is
\emph{acyclic}.

\subsection{The Generalized Hash Trie}\label{sec:ght}

\begin{figure}
\begin{lstlisting}
interface GHT {
  # fields
  relation: String, vars: Vec<String>  
  # constructor
  fn new(name: String, schema: Vec<Vec<String>>) -> Self
  # methods
  fn iter() -> Iterator<Tuple>
  fn get(key: Tuple) -> Option<GHT> }
\end{lstlisting}
\caption{The \GHT interface.}
\label{fig:ght}
\end{figure}

To unify binary join and \GJ, 
  we first need to unify the data structures they work over.
We propose the Generalized Hash Trie 
  which generalizes both the hash table used in binary join 
  and the hash trie used in \GJ.

\begin{definition}[Generalized Hash Trie (\GHT)]
  A \GHT is a tree where each leaf is a vector of tuples, and
  each internal node is a hash map whose keys are tuples, and each key
  maps to a child node.
\end{definition}

We will reuse the terminology defined for tries, including
\emph{level}, \emph{node}, and \emph{leaf}, etc., for \GHTs.  We will
also use the terms \GHT and \emph{trie} interchangeably when the
context is clear.  The {\em schema} of a \GHT is the list
$[\bm y_0,\bm y_1, \ldots, \bm y_\ell]$ where $\bm y_k$ are the
attribute names of the key at level $k$. 

\begin{figure}
  \centering
  \begin{subfigure}[c]{0.4\linewidth}
  \includegraphics[width=\textwidth]{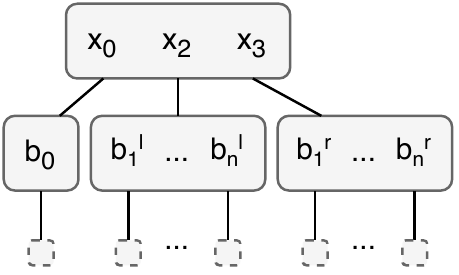}
  \end{subfigure}\hspace{1em}%
  \begin{subfigure}[c]{0.4\linewidth}
  \includegraphics[width=\textwidth]{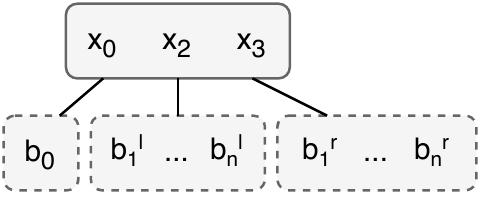}
  \end{subfigure} 
  \caption{
    Two \GHTs. The one on the left is also a hash trie, 
      and the one on the right is similar to a hash table.
    Each box with solid border stores hash keys, 
    and each box with dashed border is a vector of tuples.
    An empty box is an empty vector, representing a leaf.
  }
  \label{fig:ght-examples}
\end{figure}

The hash trie used in \GJ is a \GHT where each key is a tuple of size one,
  and the last level stores empty vectors, each of which represents a leaf.
The hash table used in binary join is very similar to a \GHT with only two levels,
  where level 0 stores the keys and level 1 stores
  vectors of tuples.
A small difference is that, in the \GHT, the concatenation of 
  a tuple from level 0 with a tuple from level 1 forms a tuple in the relation, 
  whereas each whole tuple is stored directly in a hash table.
We will show in Section~\ref{sec:colt} how the \COLT data structure
  more faithfully captures the structure of a hash table.
Figure~\ref{fig:ght-examples} shows two examples of \GHTs.

We use \GHTs to represent relations,
  and attach metadata as well as access methods 
  to each \GHT, to be used by the \FJ algorithm.
The \GHT interface is shown in Figure~\ref{fig:ght}.
The \lstinline|relation| field stores the relation name.
A sub-trie inherits its name from its parent.
The \lstinline|vars| field stores parts of the relation's schema:
  if the trie is a vector of tuples, 
  \lstinline|vars| is the schema of each tuple;
  if the trie is a map, 
  \lstinline|vars| is the schema of each key.
The constructor method \lstinline|new| creates a new \GHT from the named relation, 
  where an $n$-th level trie has variables
  matching the $n$-th element of the \lstinline|schema| argument,
  and the values along each path from the root to a leaf of the \GHT 
  form a tuple in the relation.

\begin{example}
  Both \GHTs in Figure~\ref{fig:ght-examples} represent relation $S$ from the
  clover query $Q_\clubsuit$ in Figure~\ref{fig:clover-query}.  The
  \GHT on the left (a hash trie) was created by calling the
  constructor method \lstinline|new| with the schema
  \lstinline|[[x],[b],[]]|, so the top-level trie has the schema
  \lstinline|[x]|, each second-level trie has the schema
  \lstinline|[b]|, and each third-level trie (a leaf) has the empty
  schema \lstinline|[]|.  The \GHT on the right (a hash table) was
  created by calling \lstinline|new| with the schema
  \lstinline|[[x],[b]]|.  It has only two levels, with schema
  \lstinline|[x]| and \lstinline|[b]|, respectively.  Note that each
  $b$ value in the hash trie is hashed and stored as a key, while the
  $b$ values in the hash table are simply stored in vectors.
\end{example}

The methods \lstinline|iter| and \lstinline|get| 
  provide access to values stored in the trie.
If the trie is a map, 
  \lstinline|get(key)| returns the sub-trie mapped to \lstinline|key|,
  if any.
Calling \texttt{get} on a vector returns \lstinline|None|.
If the trie is a vector, 
  \lstinline|iter()|
  returns an iterator over the tuples in the vector;
  calling \texttt{iter} on a map 
  returns an iterator over the keys.

\begin{example}
  On the second \GHT in Figure~\ref{fig:ght-examples},
    calling \texttt{iter} returns an 
    iterator over the values $[x_0, x_2, x_3]$.
    Calling \texttt{get} with the key $x_2$ 
    returns the sub-trie which is the vector $[b_1^l, \ldots, b_n^l]$.
  Calling \texttt{iter} on this sub-trie
    returns an iterator over $[b_1^l, \ldots, b_n^l]$.
\end{example}

\subsection{The \FJ plan}\label{sec:fj-plan}


A \FJ plan specifies how the \FJ algorithm should be executed.
It generalizes and unifies binary join plans and \GJ plans. 
Recall that a left-deep linear plan for binary join
  is a sequence of relations;
  it need not specify the join attributes, 
  since all shared attributes are joined.
In contrast, a \GJ plan is a sequence of variables;
  it need not specify the relations, 
  since all relations on each variable are joined.
A \FJ plan may join on any number of variables and relations at each step,
  and therefore needs to specify both explicitly.

  To help define the \FJ plan, we introduce two new concepts, called
  \emph{subatom} and \emph{partitioning}.  Fix the query $Q$ in
  Eq.~\eqref{eq:cq}:

\begin{definition}
  A \emph{subatom} of an atom $R_i(\bm x_i)$ is an expression
  $R_i(\bm y)$ where $\bm y$ is a subset of the variables $\bm x_i$.
  A \emph{partitioning} of the atom $R_i(\bm x_i)$ is a set of
  subatoms $R_i(\bm y_1), R_i(\bm y_2), \ldots$ such that
  $\bm y_1, \bm y_2, \ldots$ are a partition of $\bm x_i$.
\end{definition}

We now define the \FJ plan using these concepts.

\begin{definition}[\FJ Plan]
  Fix a conjunctive query $Q$.  A \FJ \emph{plan} is a list
  $[\phi_1, \ldots, \phi_m]$, where each $\phi_k$ is a list of
  subatoms of $Q$, called a {\em node}.  The nodes are required to
  {\em partition the query}, in the sense that, for every atom
  $R_i(\bm x_i)$ in the query, the set of all its subatoms occurring
  in all nodes must form a partitioning of $R_i(\bm x_i)$.  We denote
  by $vs(\phi_k)$ the set of variables in all subatoms of $\phi_k$.
  The variables \emph{available to} $\phi_k$ are all variables of the
  preceding nodes:
$$avs(\phi_k) = \bigcup_{j < k} vs(\phi_j)$$
\end{definition}


We will define shortly a {\em valid plan}, but first we show an example.


\begin{example}\label{ex:fj-plan}
  The following is an \FJ plan for $Q_\clubsuit$:
\begin{align}
&  [[R(x, a), S(x)], [S(b), T(x)], [T(c)]]\label{eq:bj-plan}
\end{align}
To execute the first node we iterate over each tuple $(x, a)$ in $R$
and use $x$ to probe into $S$; for each successful probe we execute
the second node: we iterate over each $b$ in $S[x]$, then use
$x$ to probe into $T$; finally the third node iterates over $c$ in
$T[x]$.  The reader may notice that this corresponds precisely to the
left-deep plan $(R(x,a) \Join S(x,b))\Join T(x,c)$.
Another \FJ plan for $Q_\clubsuit$ is: 
\begin{align}
  [[R(x), S(x), T(x)], [R(a)], [S(b)], [T(c)]]\label{eq:gj-plan}
\end{align}
This plan corresponds to the \GJ plan $[x,a,b,c]$.  Intuitively, here
we start by intersecting $R.x \cap S.x \cap T.x$, then, for each $x$
in the intersection, we retrieve the values of $a$, $b$, and $c$ from
$R$, $S$, and $T$, and output their Cartesian product.
\end{example}



Not all \FJ plans are valid, and only valid plans can be executed.  We
execute each \FJ node by iterating over one relation in that node, and
probe into the others.  Therefore, the values used in each probe must
be available, either from the same node or a previous one.

\begin{definition}
  A \FJ plan is \emph{valid} if for every node $\phi_k$ the following
  two properties hold.  (a) No two subatoms share the same relation,
  and (b) there is a subatom containing all variables in
  $vs(\phi_k) - avs(\phi_k)$.  We call such an subatom a \emph{cover}
  for $\phi_k$, and write $cover(\phi_k)$ for the set of covers.
\end{definition}

We will assume only valid plans in the rest of the paper.  To simplify
the presentation, in this section we assume that each node $\Phi_k$,
has {\em one} subatom designated as cover, and will always list it as
the first subatom in $\Phi_k$.  We will revisit this assumption in
Sec.~\ref{sec:optimization}, and allow for multiple covers.


\begin{example}
  Both plans in Example~\ref{ex:fj-plan} are valid.  The covers for
  the 3 nodes for Eq.~\eqref{eq:bj-plan} are $R(x, a)$, $S(b)$, and
  $T(c)$, respectively. For the plan in Eq.~\eqref{eq:gj-plan}, the
  covers for the 4 nodes are $R(x), R(a), S(b), T(c)$; for the first
  node we could have also chosen $S(x)$ or $T(x)$ as cover.
\end{example}
 

\begin{example}
  An example of an \emph{invalid} plan for the clover query has one
  single node containing all relations and variables:
  $$[[R(x, a), S(x, b), T(x, c)]]$$
  Intuitively, we cannot execute it: if we iterate over, say $R$, then
  we bind two variables $x$ and $a$, but to lookup $S$ we
  need the key $(x,b)$.
\end{example}






\subsection{Execution of the \FJ Plan}\label{sec:free-join-algorithm}


The execution of a \FJ plan has two phases: the build phase and the
join phase.  The build phase constructs the \GHTs for the relations in
the query, by calling the constructor method \lstinline|new| on each
relation with the appropriate schema.  The join phase works over the
\GHTs to compute the join of the relations.


\begin{figure}
  \begin{lstlisting}[
    numbers=left,
    numberstyle=\small\color{gray}\ttfamily,
    commentstyle=\color{gray},
    escapeinside={(*}{*)},
    ]
fn join(all_tries, plan, tuple):
  if plan == []:
    output(tuple) (* \label{lst:output} *)
  else:
    tries = [ t $\in$ all_tries | t.relation $\in$ plan[0] ] (* \label{lst:select} *)
    # iterate over the cover
    @outer for t in tries[0].iter(): (* \label{lst:outer} *)
      subtries = [ iter_r.get(t) ] (* \label{lst:inner} *)
      tup = tuple + t
      # probe into other tries
      for trie in tries[1..]:     
        key = tup[trie.vars] (* \label{lst:key} *)
        subtrie = trie.get(key)
        if subtrie == None: continue @outer
        subtries.push(subtrie) (* \label{lst:inner-end} *)
      new_tries = all_tries[tries $\mapsto$ subtries] (* \label{lst:tuple} *)
      join(new_tries, plan[1:], tup) (* \label{lst:join} *)
\end{lstlisting}
  \caption{The \FJ algorithm.}
  \label{fig:fj-algo}
\end{figure}

\subsubsection*{Build Phase}
The build phase constructs a \GHT for each relation (atom)
$R_i(\bm x_i)$, as follows.  If the plan partitions the atom into the
subatoms $R_i(\bm y_0), R_i(\bm y_1), \ldots, R_i(\bm y_{\ell-1})$,
then the schema of its \GHT is the list
$[\bm y_0, \bm y_1, \ldots, \bm y_{\ell-1}, []]$.  Recall that the
last level of a \GHT is a vector instead of a hash map. As an
optimization, if the last subatom $R_i(\bm y_{\ell-1})$ is the cover
of its node, then we drop the last $[]$ from the schema, in other
words, we construct a vector for the $\bm y_{\ell-1}$.  After
computing the schema for each relation, we call the constructor method
\lstinline|new| on each relation and its computed schema to build the
\GHTs.

\begin{example}
  Consider the plan in Eq.~\eqref{eq:bj-plan} for the clover query
  $Q_\clubsuit$.  The \GHT schemas for $R$, $S$, and $T$ are
  $[[x, a]]$, $[[x], [b]]$, and $[[x], [c]]$ respectively.  Thus, $R$
  is a flat vector of tuples, and each of $S$ and $T$ is a hash map of
  vectors of values.  Consider now the triangle query $Q_{\triangle}$
  and the plan $[[R(x,y),S(y),T(x)],[S(z),T(z)]]$.  The \GHT
  schemas for $R, S, T$ are $[[x,y]]$, $[[y],[z]]$, and $[[x],[z],[]]$:
  in other words $R$ is stored as a vector, $S$ is a hash-map of
  vectors, and $T$ is a hash-map of hash-maps of vectors.
\end{example}

\subsubsection*{Join Phase}
The pseudo-code for the \FJ algorithm is shown in Figure~\ref{fig:fj-algo}.
The \lstinline|join| method takes as input the \GHTs, the \FJ plan, 
  and the current tuple initialized to be empty.
If the plan is empty, we output the tuple (line~\ref{lst:output}).
Otherwise, we work on the first node in the plan
  and intersect relevant tries (line~\ref{lst:select}).
We iterate over tuples in the covering relation, 
  which is the first trie in the node (line~\ref{lst:outer}).
%
%
Then, we use values from \lstinline|t| 
  and the \lstinline|tuple| argument
  as keys to probe into the other tries 
  (line~\ref{lst:inner}-\ref{lst:inner-end}).
To construct a key for a certain trie, 
  we find the values mapped from the trie's schema variables
  in \lstinline|t| and \lstinline|tuple| (line~\ref{lst:key}).
If any probe fails, we continue to the next tuple in the outer loop.
If all probes succeed, we replace the original tries with the 
  subtries returned by the probes,
  and recursively call \lstinline|join| 
  on the new tries and the rest of the plan (line~\ref{lst:tuple}-\ref{lst:join}).

The recursive definition may obscure the essence of the \FJ algorithm,
  so we provide some examples where we unroll the recursion.
We introduce some convenient syntax to simplify the presentation.
We write \lstinline|for (x,y,...) in T:| 
  to introduce a for-loop iterating over \lstinline|T|, 
  binding the values of each tuple in \lstinline|T.iter()| 
  to the variables \lstinline|x,y,...|.
We write \lstinline|r = R[t]?| to bind the result of 
  \lstinline|R.get(t)| to \lstinline|r|;
  if the lookup fails, we continue to the next iteration of the 
  enclosing loop.
In other words, \texttt{r = R[t]?} is equivalent to:
\begin{lstlisting}
r = R.get(t); if r.is_none(): continue
\end{lstlisting}

\begin{figure*}
  \begin{subfigure}[b]{0.3\linewidth}
\begin{lstlisting}[escapeinside={(*}{*)}]
R = GHT("R",[["x","a"]])
S = GHT("S",[["x"],["b"]])
T = GHT("T",[["x"],["c"]])
for (x, a) in R:
  s = S[x]?
  for b in s:
   (* \underline{t = T[x]?} *)
    for c in t:
      output(x, a, b, c)
\end{lstlisting}
    \caption{Binary \FJ.}
    \label{fig:bj-loop}
  \end{subfigure}
  \begin{subfigure}[b]{0.3\linewidth}
    \centering
\begin{lstlisting}[
    escapeinside={(*}{*)}
]
# same as the left
# ...
# ...
for (x, a) in R:
  s = S[x]?
 (* \underline{t = T[x]?} *)
  for b in s:
    for c in t:
      output(x, a, b, c)
\end{lstlisting}
    \caption{Factorized \FJ.}
    \label{fig:factorized-loop}
  \end{subfigure}
  \begin{subfigure}[b]{0.3\linewidth}
    \centering
\begin{lstlisting}
R = GHT("R",[["x"],["a"]])
# same as the left
# ...
for x in R:
  r=R[x]?; s=S[x]?; t=T[x]?
  for a in r:
    for b in s:
      for c in t:
        output(x, a, b, c)
\end{lstlisting}
    \caption{Generic \FJ.}
    \label{fig:gj-loop}
  \end{subfigure}
  \Description[TODO]{TODO}
  \caption{Execution of \FJ for the clover query.}
\end{figure*}

\begin{example}\label{ex:binary-free-join}
  Consider the plan in Eq.~\eqref{eq:bj-plan} for the clover query $Q_\clubsuit$.
  Figure~\ref{fig:bj-loop} shows its execution; ignore the underlined
  instruction for now.
  In the build phase, 
    we construct a flat vector for $R$ and a hash table for each of $S$ and $T$.
  In the join phase, for the  node $[R(x, a), S(x)]$  we 
   iterate over $R$ and probe into $S$, while for the second node $[S(b), T(x)]$,
    we iterate over the second level of $S$ and probe into $T$.
  Finally, the third loop iterates over the second level of $T$ and outputs the result.
\end{example}

\begin{example}\label{ex:generic-free-join}
  Consider now the plan in Eq.~\eqref{eq:gj-plan} for  $Q_\clubsuit$.
  Its execution is shown in Figure~\ref{fig:gj-loop}.
  We construct  hash tables for $R$, $S$, and $T$, keyed on $x$.
  The first loop level intersects the three relations on $x$, 
    and subsequent loop levels take the Cartesian product of the relations on $a$, $b$, and $c$.
\end{example}

Note that Fig.~\ref{fig:bj-loop}
  follows the execution of binary hash join with the plan $[R, S, T]$,
  whereas Fig.~\ref{fig:gj-loop} follows the execution of \GJ
   with the plan $[x, a, b, c]$.  We will describe
   Fig.~\ref{fig:factorized-loop} later.

\subsection{Discussion}

\FJ plans generalize both traditional binary plans and \GJ.  
One
limitation so far is our assumption that the cover is chosen during
the {\em build phase}.  This was convenient for us to illustrate how
to avoid constructing some hash maps, by storing the last level of a
\GHT as vector, when it corresponds to a cover.  In contrast, \GJ
computes the intersection $R_1.x\cap R_2.x\cap \cdots$ by iterating
over the smallest set, hence it chooses the ``cover'' at run time.  We
will address this in the next section by describing \COLT, a data
structure that constructs the \GHT lazily, at run time, allowing us to
choose the cover during the {\em join phase}.

\section{Optimizing the \FJ Plan}

\label{sec:optimization}

In the previous section we have introduced \FJ plans and their associated
data structures, the \GHTs.  We have seen that a \FJ plan is capable
of covering the entire design space in Fig.~\ref{fig:design-space},
from traditional join plans to \GJ.  In this section we describe how
to build, optimize, and speedup the execution of a \FJ plan.  We start
from a conventional binary plan produced by a query optimizer, and convert
it into an optimized \FJ plan (Section~\ref{sec:bj-to-fj}). Next, we
introduce the \COLT data structure to greatly reduce the cost of
building the hash tries (Section~\ref{sec:colt}).  We present a simple
vectorized execution algorithm for \FJ
(Section~\ref{sec:vectorized-execution}), and finally, we discuss how
\FJ relates to \GJ (Section~\ref{sec:fj-gj-multijoin}).


\subsection{Building and Optimizing a \FJ Plan}\label{sec:bj-to-fj}

Our system starts from an optimized binary plan produced by a
traditional cost-based optimizer; in particular, we use DuckDB's
optimizer~\cite{DBLP:conf/cidr/RaasveldtM20,DBLP:conf/vldb/Raasveldt22}. We
decompose a bushy plan into a set of left-deep plans, as described in
Sec.~\ref{sec:background}, then convert each left-deep plan into an
equivalent \FJ plan.  Finally, we optimize the converted \FJ plan,
resulting in a plan that can be anywhere between a left-deep plan or a
\GJ plan.


\begin{figure}
  \begin{lstlisting}
fn binary2fj(bin_plan):
  fj_plan = []; r = bin_plan[0]
  $\phi_0$ = [ r(r.schema) ]; $\phi$ = $\phi_0$ # iterate over left relation
  for s in bin_plan[1:]:
    $\phi$.push(s(s.schema $\cap$ $avs(\phi)$)) # probe w/ available vars
    fj_plan.push($\phi$)
    $\phi$ = [ s(s.schema - $avs(\phi)$) ] # iterate over probe result
  fj_plan.push($\phi$)
  return fj_plan
  \end{lstlisting}
  \caption{Translating a binary plans to a \FJ plan.}
  \label{fig:bj2fj}
\end{figure}

The conversion from a binary plan to an equivalent \FJ plan is done by the function
\lstinline|binary2fj| in Figure~\ref{fig:bj2fj}.  We begin by adding the full atom of the left
relation as the first subatom in the first \FJ plan node.  Then we iterate over
the remaining relations in the binary join plan.  For each relation, we add a
subatom whose variables are the intersection of the relation's schema with the
available variables at the current \FJ plan node. Then we create a new join
node, adding to it the relation with the remaining variables.  

\begin{example}
  The binary plan $[R, S, T]$ for the clover query $Q_\clubsuit$ is
  converted into the \FJ plan shown in Eq.~\eqref{eq:bj-plan}.  For
  another example, consider a chain query:
$$Q \cd R(x, y), S(y, z), T(z, u), W(u, v).$$
The left-deep plan $[R, S, T, W]$ is converted into:
$$[[R(x, y), S(y)], [S(z), T(z)], [T(u), W(u)], [W(v)]]$$
\end{example}

So far the algorithm in Figure~\ref{fig:bj2fj} produces a \FJ plan
that is equivalent to the left-deep plan.  Next, we
optimize the \FJ plan. The main idea behind our optimization is to
bring the query plan closer to \GJ, without sacrificing the benefits
of binary join.

For intuition, let us revisit the clover query $Q_\clubsuit$, and its
execution depicted in Fig.~\ref{fig:bj-loop} (as explained in
Example~\ref{ex:binary-free-join}).  Consider the input shown in
Fig.~\ref{fig:clover-vis}.  Both relations $R$ and $S$ are skewed on
the value $x_2$, and their join will produce $n^2$ tuples, namely
$\setof{(x_2, a_i, b_j)}{i, j \in [1..n]}$.  This means the body of
the second loop in Figure~\ref{fig:bj-loop} is executed $n^2$ times.
However, the $n^2$ tuples are only to be discarded by the join with
$T$ which does not contain $x_2$.

There is a simple fix to the inefficiency: we can pull the underlined
lookup on $T$ in Figure~\ref{fig:bj-loop} out of the loop over $s$ to
filter out redundant tuples early.  This results in the nested loops
in Figure~\ref{fig:factorized-loop} which runs in $O(n)$ time, because
the two lookups in the first loop already filter the result to a
single tuple.  At the logical level, we convert the first \FJ plan
into the second \FJ plan:
  \begin{align*}
&\mbox{Naive plan (Eq.~\eqref{eq:bj-plan}):}&& [[R(x, a), S(x)], [S(b), T(x)], [T(c)]]\\
&\mbox{Optimized plan:} && [[R(x, a), S(x), T(x)], [S(b)], [T(c)]]
  \end{align*}
  While this is closer to the \GJ in Figure~\ref{fig:gj-loop}, it
  differs in that it still uses the same \GHTs built for original
  plan, without the need for an additional hash table for $R$.

\begin{figure}
  \begin{lstlisting}
fn factor(plan):
  @outer: for i in [1..n-1].reverse():
    $\phi$ = plan[i]; $\phi'$ = plan[i-1]
    for $\alpha$ in $\phi$:
      if $\alpha$.vars $\subseteq avs(\phi) \wedge \alpha$.relation $\notin \phi':$
        $\phi$.remove($\alpha$); $\phi'$.push($\alpha$)
      else: continue @outer
  \end{lstlisting}
  \caption{Factorizing a \FJ plan.}
  \label{fig:factorize-plan}
\end{figure}

More generally, we will optimize a \FJ plan by \emph{factoring out}
lookups, i.e. by moving a subatom from a node $\Phi_i$ to the node
$\Phi_{i-1}$.  In doing so we must ensure that the plan is still
valid, and also avoid accidental slowdowns.  For example, we cannot
factor the lookups on $S$ and $T$ beyond the outermost loop, because
that loop binds the variable $x$ used in the lookups.

The optimization algorithm for \FJ plans is shown in
Figure~\ref{fig:factorize-plan}. We traverse the plan in reverse order
visiting each node. For each node, if there is an atom whose variables
are all available before that node, and if the previous node does not
contain an atom of the same relation, we move the atom to the previous
node. These two checks ensure the factored plan remains valid.  The
last line in the algorithm ensures we factor lookups
\emph{conservatively}. That is, we factor out a lookup only if all
previous lookups in the same node have also been factored out. Doing
so respects the lookup ordering given by the original cost-based
optimizer, since scrambling this ordering may inadvertently slow down
the query. It should be clear that, except for extreme cases where the
enclosing loop is empty, factoring out any lookup will always improve
performance.


\subsection{\COLT: Column-Oriented Lazy Trie}\label{sec:colt}

The original \GJ algorithm builds a hash trie for each input relation.
A left-deep plan avoids building a hash table on the left most
relation, since it only needs to iterate over it, and this is an
important optimization, since the left most relation is often the
largest one.  Building a subtrie can also be wasteful when that
subtrie's parent is pruned away by an earlier join, in which case the
subtrie will never be used.  To address that, we describe here how to
build the tries \emph{lazily}: we only build the trie for a
(sub-)relation at runtime, if and when we need to perform a lookup, or
need to iterate over a prefix of its tuples.  This idea leads to our
new data structure called Column-Oriented Lazy Trie, or \COLT for
short.  In our system  the raw data is stored column-wise, in main
memory, and each column is stored as a vector, as standard in
column-oriented databases~\cite{DBLP:journals/ftdb/AbadiBHIM13}.
\begin{definition}
  A \emph{\COLT} is a tree where each leaf is a vector of offsets into
  the base relation, and each internal node is a hash map mapping a
  tuple to a child node.
\end{definition}

\begin{figure}
  \includegraphics[width=0.4\linewidth]{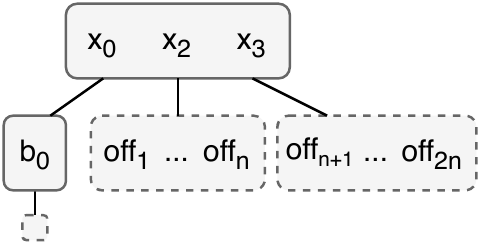}
  \caption{A \COLT for the relation $S$ in
    Fig.~\ref{fig:clover-query}. Each off$_i$ is an integer
    representing an offset into the base table $S$.}
  \label{fig:colt}
\end{figure}

A \COLT tree need not be balanced, and there can be both hash maps and
vectors at the same tree level.  Fig.~\ref{fig:colt}
illustrates a \COLT tree for the instance $S$ of the clover query
$Q_\clubsuit$.


\begin{figure}
  \begin{lstlisting}
struct COLT {
  relation, schema, vars, 
  data = Map(HashMap<Tuple, COLT>) | Vec<Vec<u64>> }

impl GHT for COLT:
  fn new(relation, schema):
    COLT { relation, schema, schema[0],
           data = [ 0, 1, ..., relation.len - 1 ] }

  fn iter():
    match self.data:
      Map(m) => m.keys().iter(),
      Vec(v) => 
          if is_suffix(self.vars, relation.schema):
            v.map(|i| cols = self.relation[self.vars];
                      cols.map(|c| c[i]) )
          else: self.force(); self.iter()

  fn get(key): self.force(); self.get_map.get(key)

  fn force():
    match self.data:
      Map(m) => {} # already forced, do nothing
      Vec(v) => 
        map = new()
        for i in v:
          cols = self.relation[self.vars]
          k = cols.map(|col| col[i])
          if map[k] is None: # make a new, empty COLT
            map[k] = COLT { relation: self.relation, 
                            schema: self.schema[1..], 
                            data: [] }
          map[k].data.push(i)
        self.data = Map(map)
  \end{lstlisting}
  \caption{The \COLT data structure.}
  \label{fig:colt-impl}
\end{figure}


\COLT Implements the \GHT interface in Figure~\ref{fig:ght}, and its
implementation is shown in Figure~\ref{fig:colt-impl}. As before,
\COLT stores a reference to the relation it represents, as well as the
\GHT schema computed from the plan.
Consider a relation with $n$ tuples.  The \COLT tree is initialized
with a single node consisting of the vector $[0, \ldots, n-1]$,
i.e. one offset to each tuple.  
\COLT
implements the \lstinline|get| and \lstinline|iter| methods lazily.  
When
\lstinline|get| is called, we check if the current node is a hash
map or a vector.  In the first case, we simply perform a lookup in the
map.  In the second case, we first replace the current vector with a
hash map, whose children are vectors of offsets.  Notice that this
requires iterating over the current vector of offsets, accessing the
tuple in the base table, inserting the key in the hash map, and
inserting the offset in the corresponding child.  Consider now a call
to \lstinline|iter|.  If the current node is a hash map, then we
return an iterator over it.  If it is a vector, then we check if it is
a suffix of the relation schema: if yes, then we simply iterate over
that vector (and access the tuples via their offsets), otherwise we
first materialize the current hash map as explained above, and return
an iterator over the hash map.

As a simple but effective optimization, we do not initialize the \COLT
tree to the single node $[0,1,\ldots,n-1]$, but instead iterate
directly over the base table, if required.  If no \lstinline|get|
is performed on this table, then we have completely eliminated the
cost of building any auxiliary structure on this table.  Thus, the \FJ
plan can be equivalent to a left-deep plan that avoids building a hash
table on the left-most relation.  \COLT is also closer to the structure of
traditional hash tables, which, in some implementations, map a key to
a vector of pointers to tuples.
%
%

\begin{example}\label{ex:colt-clover}
Consider an extension of the clover query $Q_\clubsuit$:
\begin{align*}
  Q(x, a, b, c) \cd R(x,a), S(x,b), T(x, c), \underline{U(b)}.
\end{align*}
  %
%
\GJ builds a 2-level hash trie for each of $R$, $S$, and $T$, as well
as a 1-level hash trie for $U$.  Consider the \FJ plan
$[[R(x,a),S(x),T(x)],[U(b),S(b)],[T(c)]]$.  \FJ executes the first
node of the plan by iterating over $R$ directly, without constructing
any auxiliary structure. For each tuple $(x, a)$ in $R$, it looks up
$x$ in $S$ and $T$.  Upon the first lookup, \COLT builds the first
level of the \GHT for $S$ and $T$, i.e. a hash map indexed by the $x$
values.  Assuming the database instance for $R,S,T$ shown in
Fig.~\ref{fig:clover-query}, the result of $R.x \cap S.x \cap T.x$ has
only one value, $x_0$, thus, $\FJ$ executes the second node for only
one value $x_0$.  Here it needs to intersect $U(b)$ and $S(b)$.
Assume for the moment that \FJ chooses $U(b)$ to be the cover, on the
first lookup in $S$, \COLT will expand the second level, arriving at
Figure~\ref{fig:colt}: notice that all other $b$ values in $S$ will
never be inserted in the hash table.  More realistically, \FJ follows
the principle in \GJ and chooses $S(b)$ as cover, because it is the
smallest: it builds a hash map for $U$, but none for the 2nd level of
$S$.
\end{example}

The example highlights a divergence between \GJ and traditional plans.
To intersect $R_1.x \cap R_2.x \cap \ldots$, \GJ choose to iterate
over the smallest relation, which results in the best runtime {\em
  ignoring} the build time.  A traditional join plan will iterate over
the largest relation, because then it needs to build hash tables only
on the smaller relations.  Currently, we follow \GJ, and plan to
explore alternatives in the future.



\subsection{Vectorized Execution}\label{sec:vectorized-execution}
The \FJ algorithm as presented in Figure~\ref{fig:fj-algo}
  suffers from poor temporal locality.
In the body of the outer loop, 
  we probe into the same set of relations for each tuple.
However, these probes are interrupted by the recursive 
  call at the end, 
  which is itself a loop interrupted by further recursive calls.

\begin{figure}
  \begin{lstlisting}
@outer for ts in tries[0].iter_batch(batch_size):
  tup_subtries = [(tuple + t, [ tries[0].get(t) ]) | t $\in$ ts]
  for trie in tries[1..]:
    for (tup, subtries) in tup_subtries:
      subtrie = trie.lookup(tup[trie.vars])
      if subtrie is None:
        tup_subtries.remove((tup, subtries))
      else: subtries.append(subtrie)
  for (tup, subtries) in tup_subtries:
    new_tries = all_tries[tries $\mapsto$ subtries]
    join(new_tries, plan[1:], tup)
\end{lstlisting}
  \caption{Vectorized execution for \FJ.}
  \label{fig:vectorized-execution}
\end{figure}

A simple way to improve locality is to perform a batch of probes 
  before recursing, just like the classic vectorized execution 
  for binary join.
Concretely, we replace the \lstinline|iter| method 
  with a new method \lstinline|iter_batch(batch_size)|
  which returns up to \lstinline|batch_size| tuples at a time.
If there are less than \lstinline|batch_size| tuples left, 
  it returns all the remaining tuples.
Then we replace the outer loop in Figure~\ref{fig:fj-algo} 
  with the one in Figure~\ref{fig:vectorized-execution}.
For each batch of tuples, 
  we create a vector pairing each tuple to its subtrie in 
  \lstinline|tries[0]|.
Then for each trie to be probed,
  we iterate over the vector and look up each tuple 
  from the trie.
If the lookup succeeds, we append the subtrie to the vector of tries paired with the tuple.
If it fails, we remove the tuple to avoid probing it again.
Finally, with each tuple and the subtries it pairs with, 
  we recursively call \lstinline|join|.

\subsection{Discussion}\label{sec:fj-gj-multijoin}

\COLT is a lazy data structure, sharing a similar goal with database
cracking~\cite{DBLP:conf/cidr/IdreosKM07,DBLP:conf/sigmod/IdreosKM07}, where an
index is constructed incrementally, by performing a little work during each
lookup. 
Another connection is to Factorized Databases~\cite{DBLP:journals/sigmod/OlteanuS16} -- we
intentionally used the term ``factor'' when describing how we optimize \FJ plans to
suggest this connection. 
Concretely, we can view the trie data structure as a
factorized representation of a relation, where keys of the same hash map are
combined with union, and tuples are formed by taking the product of values at
different levels. Practically, we can use this factorized representation to 
compress large outputs, saving time and space during materialization.

As we discussed at the end of Section~\ref{sec:free-join}, in order
match the optimality of \GJ, the \FJ algorithm needs to choose
dynamically the ``cover'', i.e. the relation over which to iterate.  To
achieve this, we first find {\em all} covers for each node, then make a simple
change to the \FJ algorithm in Figure~\ref{fig:fj-algo}: we simply
choose to iterate over the cover whose trie has the fewest keys.  For
that we insert the following code right before the outer loop in
Figure~\ref{fig:fj-algo}:
\begin{lstlisting}
trie[0] = covers(plan[0]).min_by(|t| t.keys().len)
trie[1..] = # the rest of the tries
\end{lstlisting}
When we use \COLTs, we cannot know the exact number of keys in a vector unless
  we force it into a hash map. In that case we use the length of the vector as
  an estimate.

\begin{example}
  Consider the triangle query $Q_\triangle$, and the \FJ plan
  $[[R(x), T(x)], [R(y), S(y)], [S(z), T(z)]]$.  Each subatom is a
  cover of its own node.  On the outermost loop, we iterate over $R$
  if it has fewer $x$ values, and otherwise we iterate over $T$.  On
  the second loop level we make a decision picking between $S$ and a
  subtrie of $R$, \emph{for each subtrie of $R$}.  Finally, on the
  innermost loop we pick between the subtries of $S$ and $T$.
\end{example}

\section{Experiments}\label{sec:eval}


We implemented \FJ as a standalone Rust library.  The main entry point
of the library is a function that takes a binary join plan (produced
and optimized by DuckDB), and a set of input relations.  The system
converts the binary plan to a \FJ plan, optimizes it, then runs it
using \COLT and vectorized execution.
We compare \FJ against two baselines: our own \GJ implementation in
Rust, and the binary hash join implemented in the state-of-art
in-memory database
DuckDB~\cite{DBLP:conf/cidr/RaasveldtM20,DBLP:conf/vldb/Raasveldt22}.
We evaluate their performance on the popular Join Order Benchmark
(JOB)~\cite{DBLP:journals/pvldb/LeisGMBK015} and the LSQB
benchmark~\cite{DBLP:conf/sigmod/MhedhbiLKWS21}.  
In addition, we compare against K\`uzu~\cite{kuzu:cidr}, 
 a system that implements \GJ.
K\`uzu is the current iteration of the Graphflow system~\cite{DBLP:journals/pvldb/MhedhbiS19}.
We ask three research questions:
\begin{enumerate}
    \item How does \FJ perform compared to binary join and \GJ, on acyclic and cyclic queries?
    \item What is the impact of \COLT and vectorization on \FJ?
    \item How sensitive is \FJ to the query optimizer's quality?
\end{enumerate}


\begin{figure*}
  \centering
  \begin{minipage}[t]{.33\textwidth}
    \centering
    \includegraphics[width=\linewidth]{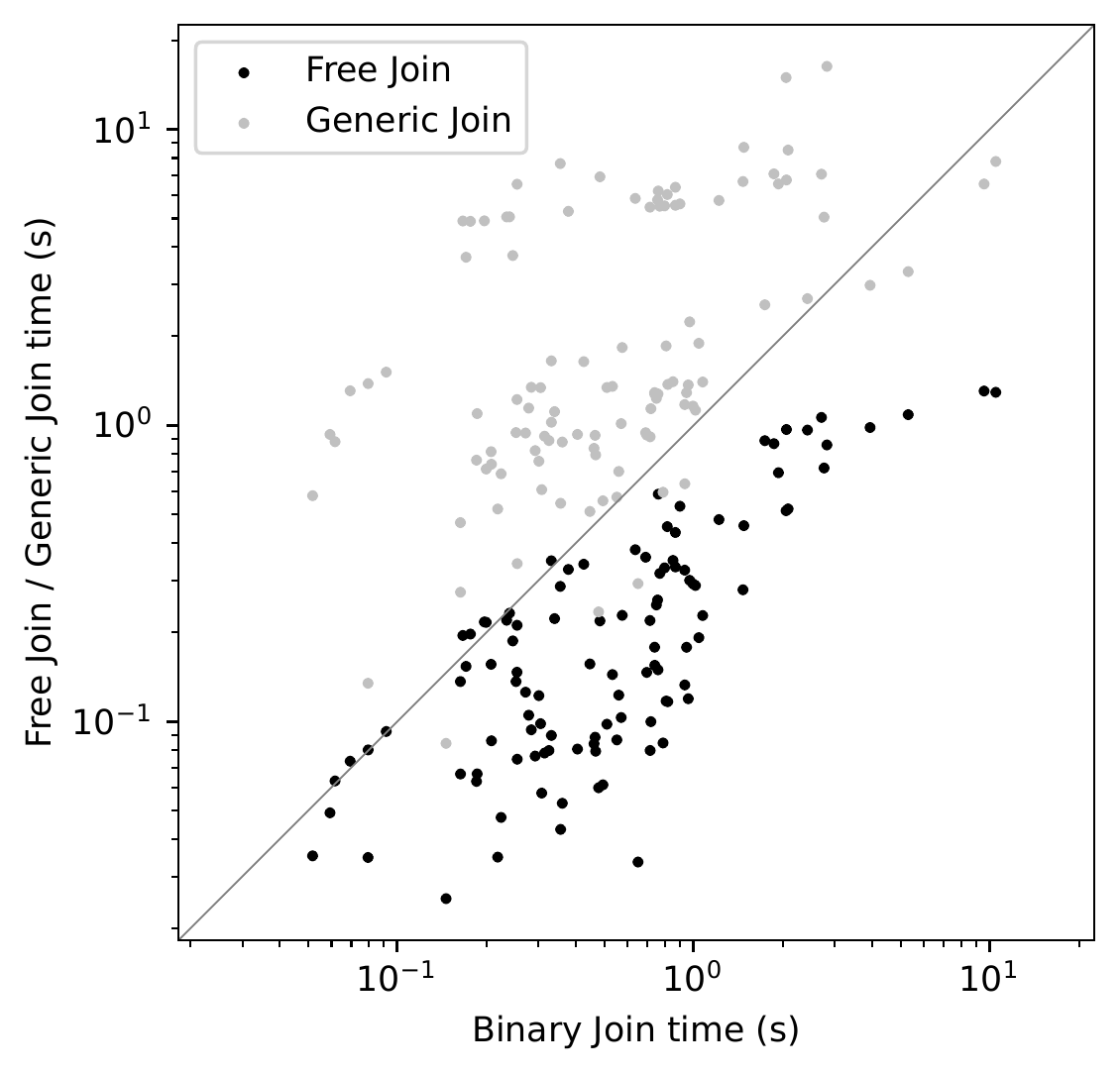}
    \captionof{figure}{Run time on JOB queries.}
    \label{fig:eval:imdb}
  \end{minipage}%
  \begin{minipage}[t]{.33\textwidth}
    \centering
    \includegraphics[width=\linewidth]{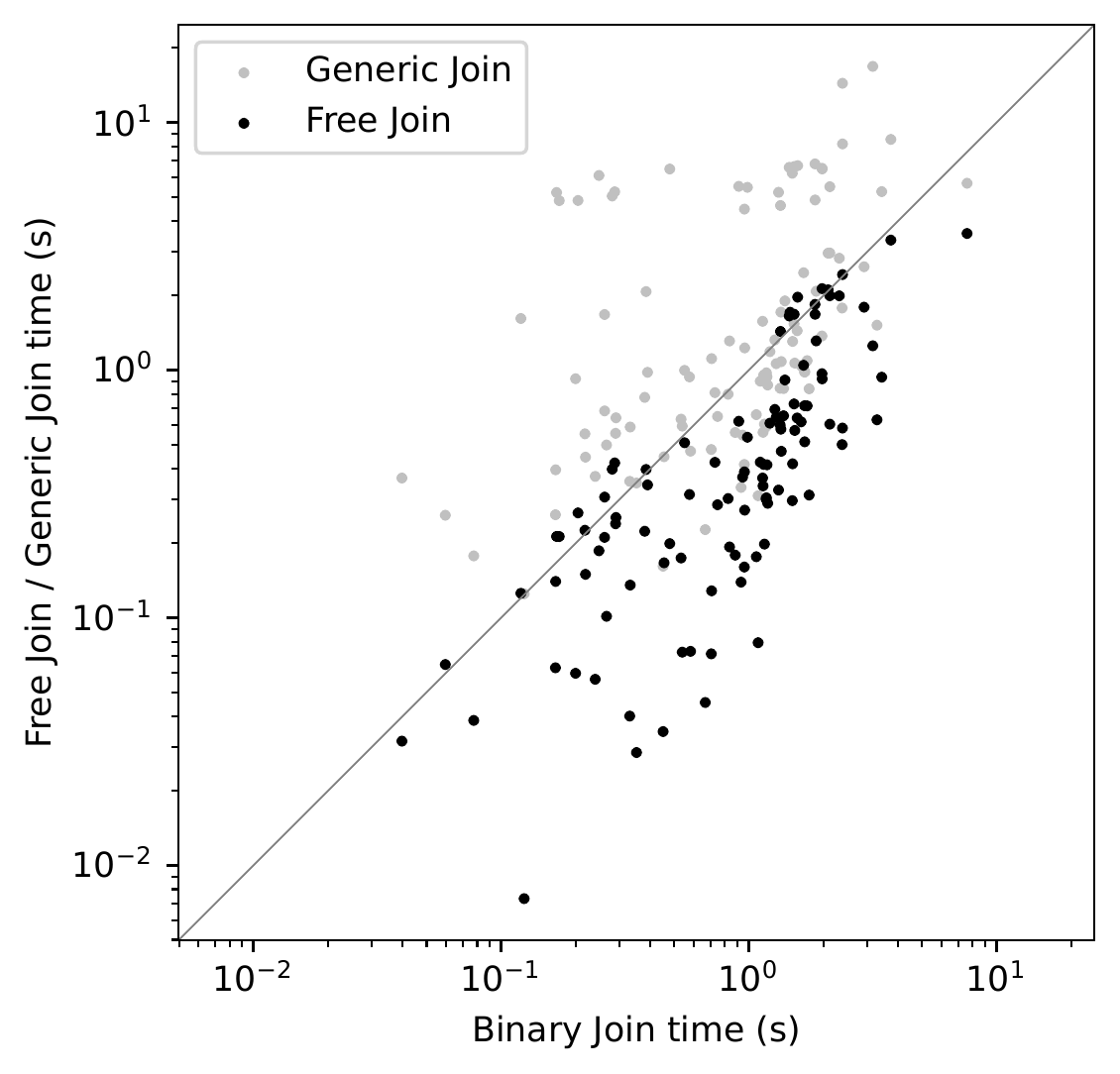}
    \captionof{figure}{Run time w/ bad cardinality estimate.}
    \label{fig:robust}
  \end{minipage}
  \begin{minipage}[t]{.33\textwidth}
    \centering
    \includegraphics[width=\linewidth]{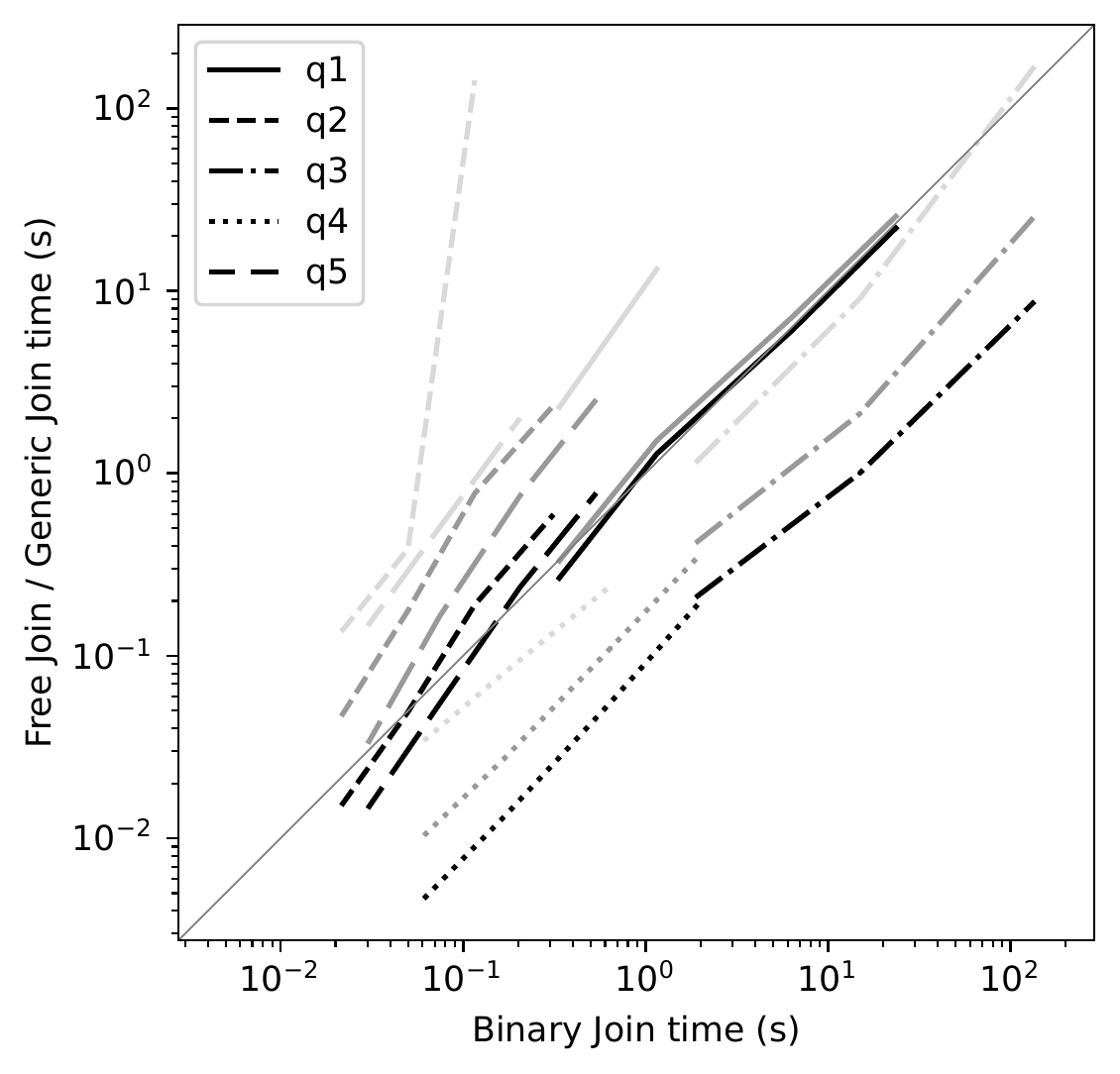}
    \captionof{figure}{Run time on LSQB queries.
    }
    \label{fig:eval:lsqb}
  \end{minipage}
  \end{figure*}
\begin{figure*}
  \centering
  \begin{minipage}{.33\textwidth}
    \centering
    \includegraphics[width=\linewidth]{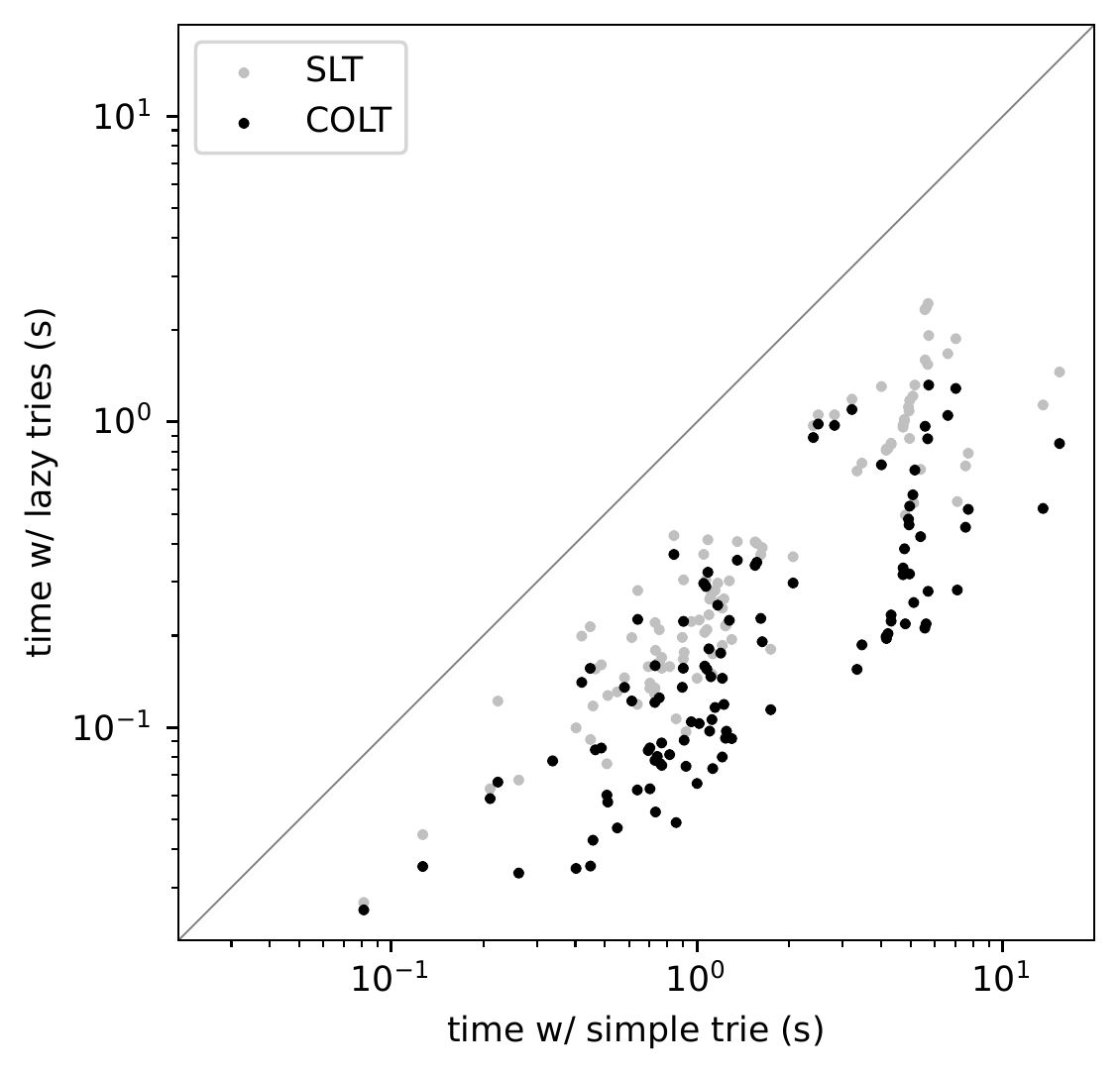}
    \captionof{figure}{Impact of \COLT.}
    \label{fig:eval:colt-ab}
  \end{minipage}%
  \begin{minipage}{.33\textwidth}
    \centering
    \includegraphics[width=\linewidth]{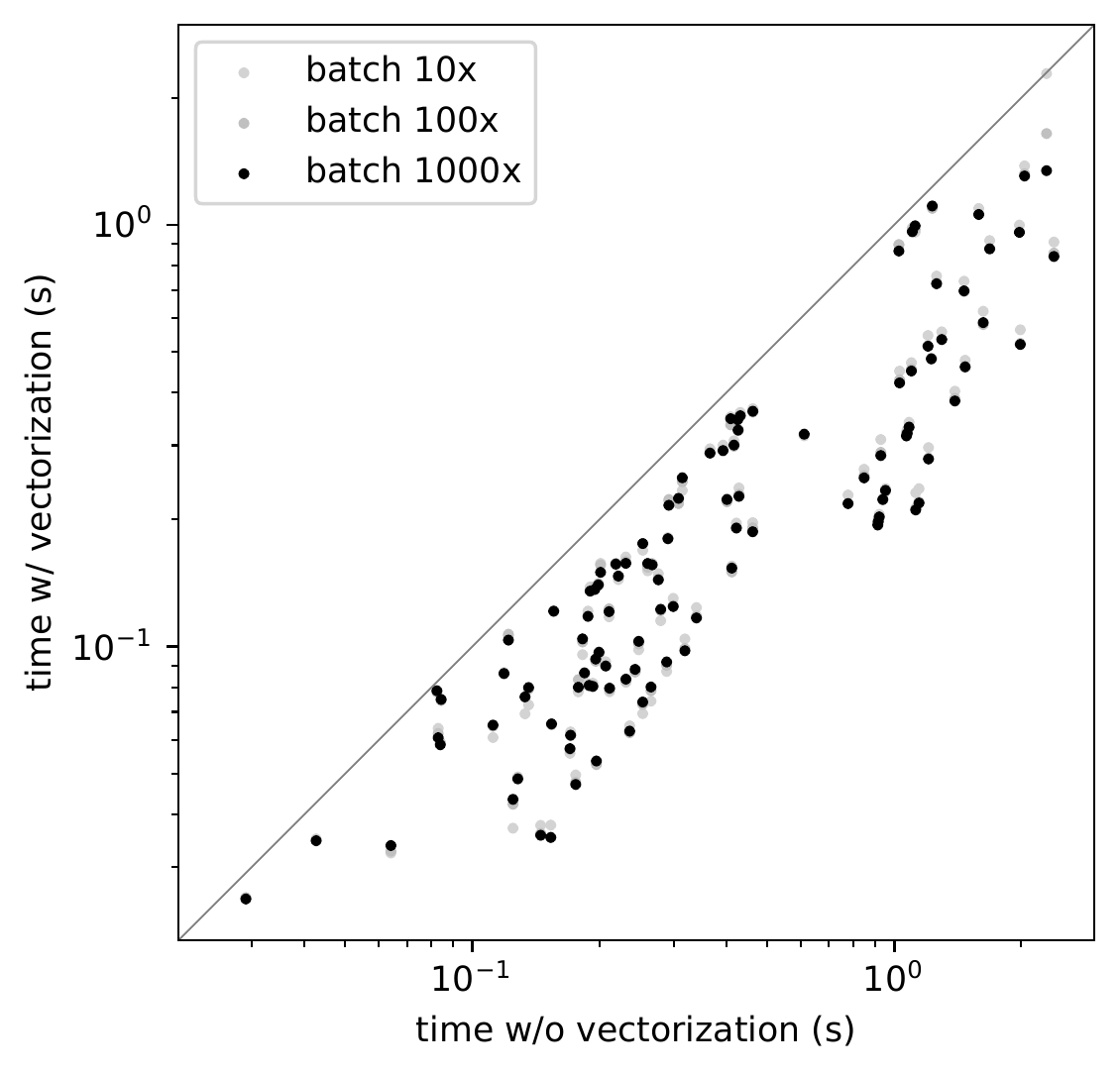}
    \captionof{figure}{Impact of vectorization.}
    \label{fig:eval:vec-ab}
  \end{minipage}
  \begin{minipage}{.33\textwidth}
    \centering
    \includegraphics[width=\linewidth]{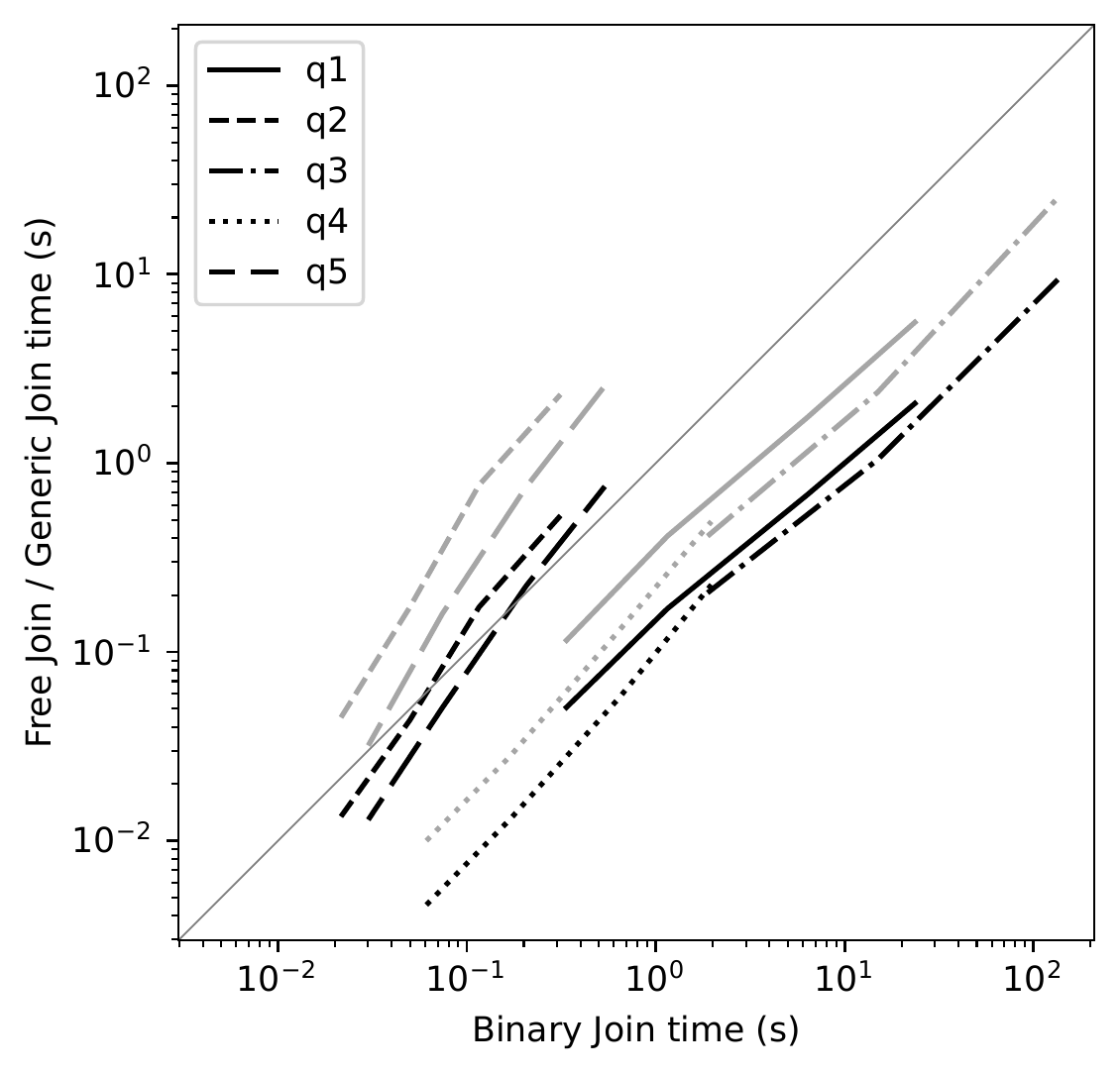}
    \captionof{figure}{LSQB run time w/ factorized output.}
    \label{fig:eval:lsqb-factor}
  \end{minipage}
  \end{figure*}
\begin{figure*}
\begin{subfigure}[t]{0.33\linewidth}
  \includegraphics[width=\linewidth]{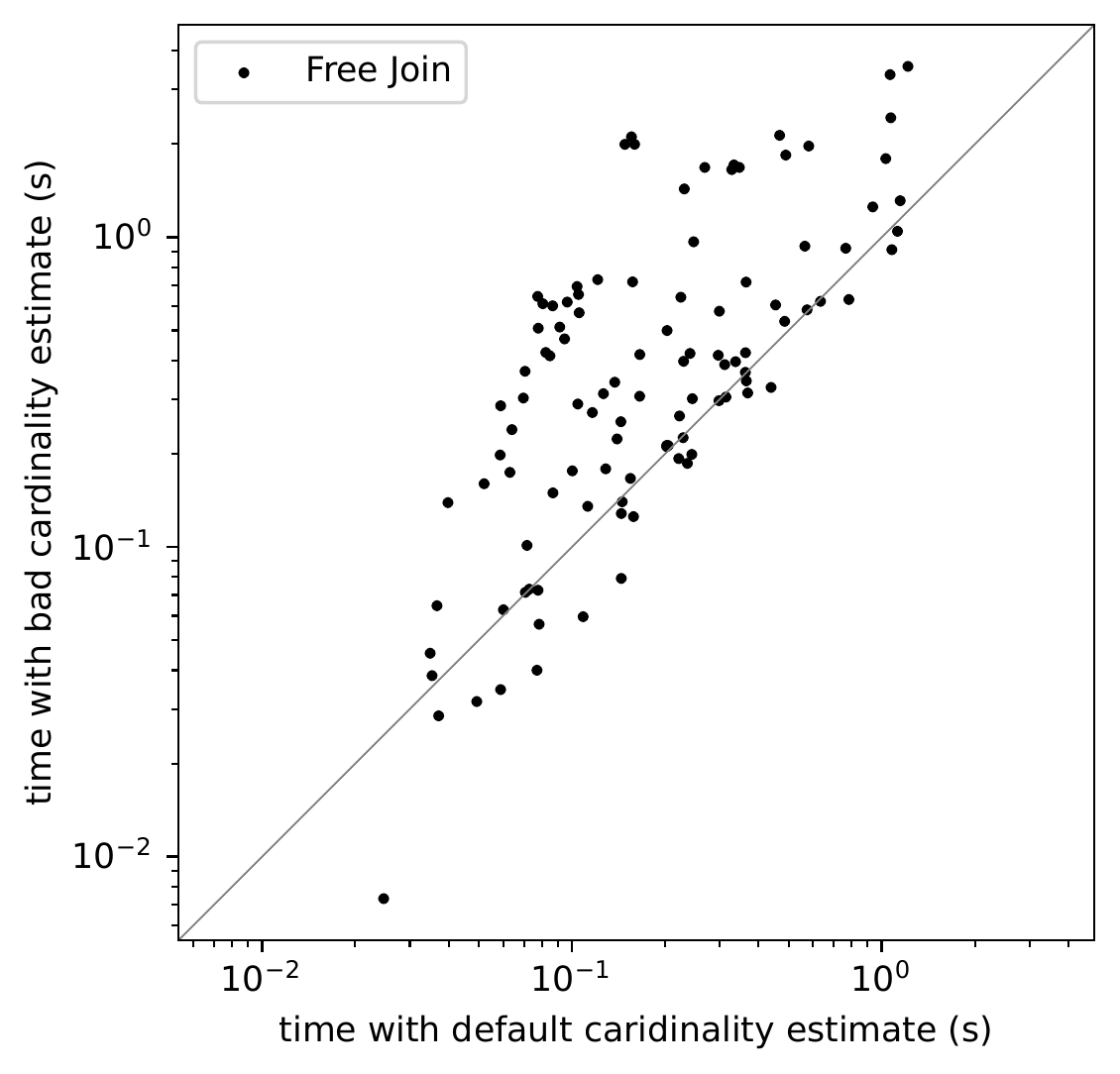}
\end{subfigure}
\begin{subfigure}[t]{0.33\linewidth}
  \includegraphics[width=\linewidth]{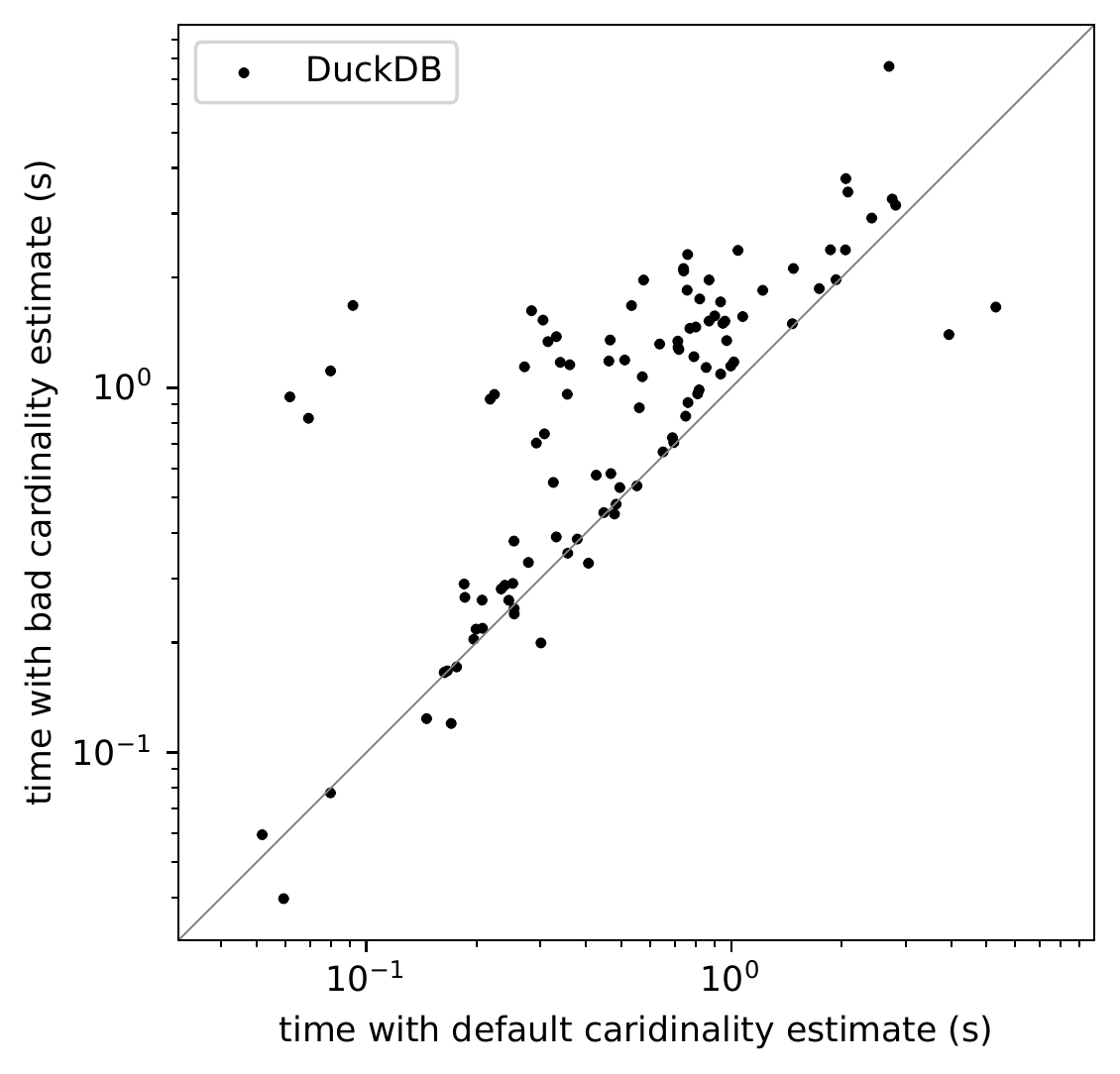}
\end{subfigure}
\begin{subfigure}[t]{0.33\linewidth}
  \includegraphics[width=\linewidth]{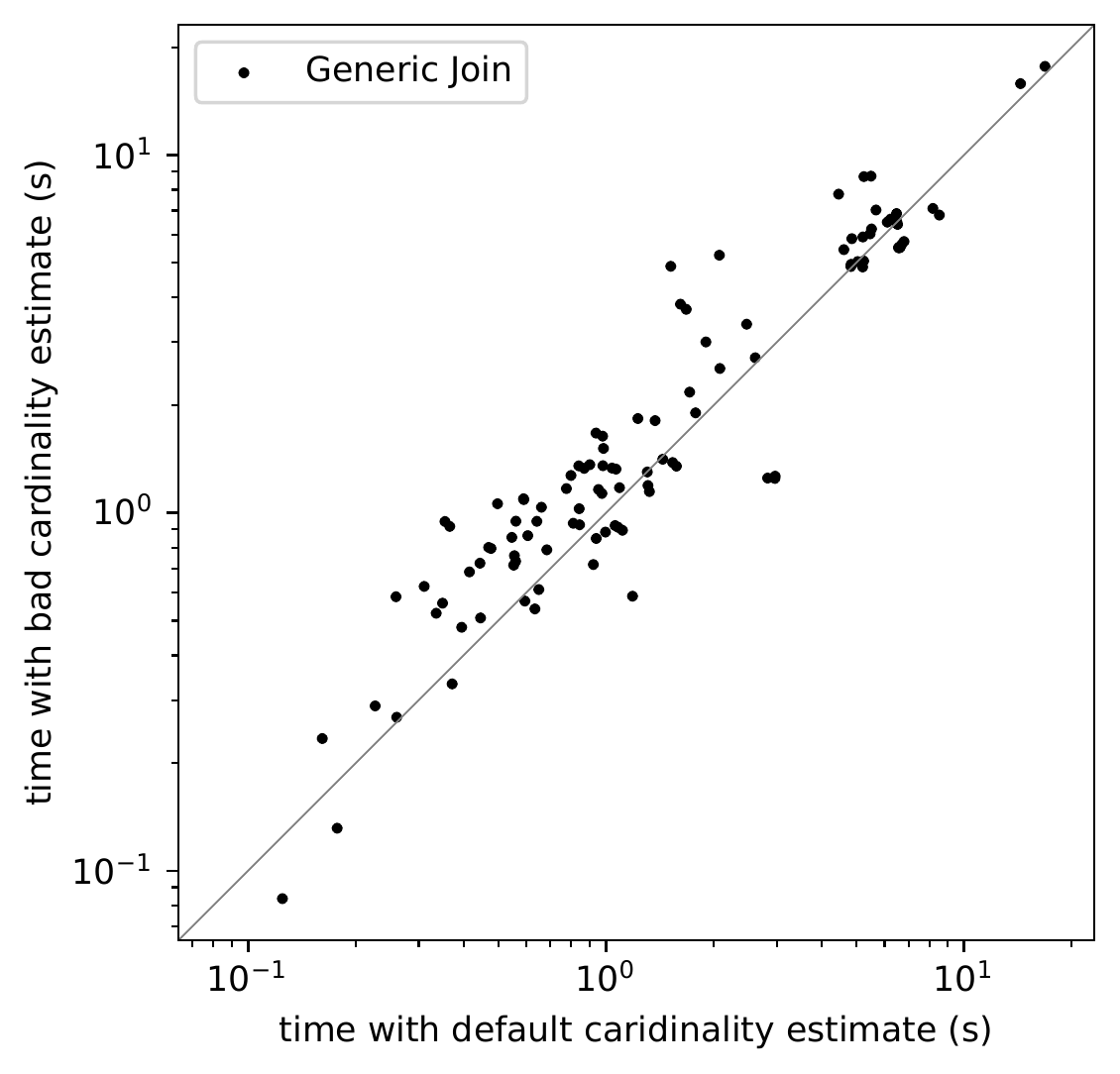}
\end{subfigure}
  \caption{Run time of join algorithms with good and bad cardinality estimates.}
  \label{fig:sensitivity}
\end{figure*}


\subsection{Setup}

While we had easy access to optimized join plans produced by DuckDB,
we did not find any system that produces optimized \GJ plans, 
or can take an optimized plan as input.
We therefore implement a \GJ baseline ourselves,
by modifying \FJ to fully construct all tries, and removing
vectorization.  We chose as variable order for \GJ the same as for
\FJ.\footnote{\FJ defines only a partial order; we extended it to a
  total order.}


Both  the JOB and the LSQB benchmarks  focus on joins.
JOB contains 113 acyclic queries with an average of 8 joins per query,
  whereas LSQB contains both cyclic and acyclic queries.
Each query in the benchmarks only contains base-table filters, 
  natural joins, and a simple group-by at the end,
  and no null values.
JOB works over real-world data from the IMDB dataset, 
  and LSQB uses synthetic data.
We exclude 5 queries from JOB that return empty results, 
  since such empty queries are known to introduce reproducibility issues\footnote{
    See GitHub issue: \url{https://github.com/gregrahn/join-order-benchmark/issues/11}
  }.
We use the first 5 queries from LSQB; the other 4 queries require 
  anti-joins or outer joins which we do not support.

We ran all our experiments on a MacBook Air laptop with Apple M1 chip and 16GB memory. 
All systems are configured to run single-threaded in main memory, 
  and we leave all of DuckDB's configurations to be the default.
All systems are given the same binary plan optimzed by DuckDB.
To answer our third research question, 
  we needed to hijack DuckDB's optimizer to produce a poor plan.
We did this by modifying its cardinality estimator to always return 1.
Since we are only interested in the performance of the join algorithm, 
  we exclude the time spent in selection and aggregation 
  when reporting performance.
This excluded time takes up on average less than 1\% of the total execution time.

\subsection{Run time comparison}\label{sec:run-time-comparison}
Our first set of experiments compare the performance of \FJ, \GJ, and binary join
  on the JOB and LSQB benchmarks.
For each query in the benchmarks, we invoke DuckDB to obtain an optimized binary plan,
  and provide the plan to our \FJ and \GJ implementation.
We run LSQB with the scaling factors 0.1, 0.3, 1, and 3, 
  as some queries run out of memory with larger scaling factors.

Figure~\ref{fig:eval:imdb} compares the run time of \FJ and \GJ against binary join on JOB queries.
We see that almost all data points for \FJ are below the diagonal, 
  indicating that \FJ is faster than binary join.
On the other hand, the data points for \GJ are largely above the diagonal, 
  indicating that \GJ is slower than both binary join and \FJ.
On average (geometric mean), \FJ is \imdbavgfjbj faster than binary join 
  and \imdbavgfjgj faster than \GJ.
The maximum speedups of \FJ against binary join and \GJ 
  are \imdbmaxfjbj and \imdbmaxfjgj, respectively, 
  while the minimum speedups are \imdbminfjbj (\imdbmaxbjfj slowdown) and \imdbminfjgj.

We zoom in onto a few interesting queries for a deeper look.
The slowest query under DuckDB is Q13a, taking over 10 seconds to finish.
  \GJ runs slightly faster, taking 7 seconds, 
  whereas \FJ takes just over 1 second.
The query plan for this query reveals the bottleneck for binary join:
  the first 3 binary joins are over 4 very large tables, 
  and two of the joins are many-to-many joins, exploding the 
  intermediate result to contain over 100 million tuples.
However, all 3 joins are on the same attribute;
in other words they are quite  similar to our clover query $Q_\clubsuit$.
As a result, \GJ and \FJ simply intersects the relations 
  on that join attribute, 
  expanding the remaining attributes only after 
  other more selective joins.
This data point appears to confirm a folklore that claims
  \WCOJ algorithms are more resistant to poor query plans.
After all, binary join could have been faster, had the query plan 
  ordered the more selective joins first.
We expand on this point with more experiments evaluating 
  each algorithm's robustness against poor plans in Section~\ref{sec:robustness}.

On a few queries \FJ runs slightly slower than binary join, 
  as shown by the data points over the diagonal.
The binary plans for these queries are all bushy, 
  and each query materializes a large intermediate relation.
We have not spent much effort optimizing for materialization, 
  and we implement a simple strategy: for each intermediate 
  that we need to materialize, we store the tuples containing 
  all base-table attributes in a simple vector.
Future work may explore more efficient materialization strategies, 
  for example only materializing attributes that 
  are needed by future joins.

Figure~\ref{fig:eval:lsqb} compares the performance of \FJ and \GJ against binary join on LSQB queries.
Each line corresponds to one query running on scaling factors 0.1, 0.3, 1, and 3.
The black lines are for \FJ, gray lines for our own \GJ baseline, and light gray lines for K\`uzu.
K\`uzu errors when loading data for SF3;
it did not finish after 10 minutes for q1 SF 1. 
DuckDB also took over 10 minutes running q3 SF 3. 
These instances do not show up in the figure.
We can see K\`uzu takes consistently longer than our \GJ implementation
 on all queries across scaling factors.
This shows that our \GJ implementation is a reasonable baseline to compare against.
On cyclic queries, \FJ is up to 15.45x (q3) faster than binary join, 
  and up to 4.08x (q2) faster than \GJ.
On acyclic queries \FJ is up to 13.07x (q4) and 3.25x (q5) faster than binary join and \GJ, respectively.
On q3 and q4 both \FJ and \GJ consistently outperform binary join on all scaling factors.
q3 contains many cycles, whereas q4 is a star query, so the superior performance of \FJ and \GJ is expected.
Surprisingly, despite q2 being a cyclic query, 
  \FJ is only slightly faster on smaller inputs
  and is even slightly slower on larger inputs.
This is the opposite of the common believe that \WCOJ algorithms
  should be faster on cyclic queries.
The query plan reveals that there are no skewed joins,
  and so binary join suffers no penalty.
Our experience shows that, in practice, 
  the superiority of \WCOJ algorithms like \FJ and \GJ
  is not solely determined by the cyclicity of the query;
  the presence of skew in the data is another important factor.

Unlike the JOB queries, in LSQB the output size (before aggregation)
  is much larger than the input size.
This leads to a large amount of time spent in constructing the output, 
  which involves random accesses to retrieve the data values for each tuple.
We therefore implemented the optimization in Section~\ref{sec:fj-gj-multijoin}
  to factorize the output.
This made q1 significant faster, as shown in Figure~\ref{fig:eval:lsqb-factor},
  while other queries are not affected.

  \subsection{Impact of \COLT and Vectorization}
The three key ingredients that make \FJ efficient 
  are \begin{enumerate*}
    \item our algorithm to optimize the \FJ plan by factoring,
    \item the \COLT data structure, and
    \item the vectorized execution algorithm.
  \end{enumerate*}
We conduct an ablation study to evaluate the performance impact of these components. 
But if we do not optimize the \FJ plan converted from a binary plan 
  and execute it as-is, \FJ would behave identically to binary join.
Since we have already compared \FJ with binary join in Section~\ref{sec:run-time-comparison},
  we do not repeat it here.
Therefore, our ablation study includes two sets of experiments, 
  evaluating the impact of \COLT and vectorization respectively.

Figure~\ref{fig:eval:colt-ab} compares the run time of \FJ using different trie data structures.
The baseline fully expands each trie ahead of time, and we call this \emph{simple trie}.
Another data structure, \emph{simple lazy trie} (SLT), expands the first level of each trie 
  ahead of time, while expanding the inner levels lazily.
This is the same strategy as proposed by~\citet{DBLP:journals/pvldb/FreitagBSKN20}.
Finally, \COLT is our column-oriented lazy trie.
In all three cases, we use the default vectorization batch size 1000.
The experiments show the average (geometric mean) speedup of \COLT 
  is 1.91x and 8.47x, over SLT and simple trie respectively,
  and the maximum speedup over them is 11.01x and 26.29x, respectively.

Figure~\ref{fig:eval:vec-ab} compares the run time of \FJ using different vectorization batch sizes.
The baseline uses no vectorization, i.e., we set the batch size to 1.
Then we adjust the batch size among 10, 100, and 1000.
The data does not show significant performance differences among
  the different batch sizes -- 
  it appears \emph{any amount of vectorization is better than none}.
For short-running queries, a smaller batch size perform slightly better,
  and for longer running queries a larger batch size wins.
We conjecture this is due to a smaller batch having less overhead, 
  leading to lower latency, 
  while a larger batch size speeds up large joins better,
  leading to better throughput. 
Overall, using the default batch size 1000
  leads to an average (geometric mean) speedup of 2.12x, 
  and a maximum speedup of 5.33x over non-vectorized \FJ.

\subsection{Robustness Against Poor Plans}\label{sec:robustness}

Our last set of experiments compares \FJ, \GJ and binary join 
  on their sensitivity to the quality of the query plan.
Many believe \WCOJ algorithms suffer less from poor query planning,
  due to its asymptotic guarantees.
Our experience with Q13 from Section~\ref{sec:run-time-comparison}
  also seems to confirm this intuition.
However, our experimental results tell a different story.
As Figure~\ref{fig:robust} shows, 
  the relative performance of the three algorithms stays 
  the same with good and bad plans, with \FJ being the fastest,
  \GJ the slowest, and binary join in the middle.
However, as shown in Figure~\ref{fig:sensitivity},
  \FJ seems to slow down as much as binary join 
  when the plan is bad (there are many points far above the diagonal).
It turns out with a poor cardinality estimate, 
  DuckDB routinely outputs bushy plans that materialize large results.
We have noted in Section~\ref{sec:run-time-comparison} that 
  our materialization strategy is simplistic, 
  so with larger intermediates it leads to more severe slowdown.
In comparison, \GJ slows down less (the data points are close to the diagonal).
However, it was the slowest to begin with,
  and since overheads like trie building dominates \GJ's run time, 
  a bad plan does not make it much slower.
Overall, we believe \FJ can be more robust to bad plans with 
  faster materialization.

\section{Limitations and Future Work}\label{sec:discussion}

With \FJ we have made a first step to bring together 
the worlds of binary join and \WCOJ algorithms.

There are three obvious limitations that require future work.  Our
current system is main memory only.  If the data were to reside on
disk, then \COLT could be quite inefficient, because it requires
repeated, random accesses to the data.  Another limitation is that the
optimization of a \FJ plan is split into two phases: a
traditional cost-based optimization (currently done by DuckDB),
followed by a heuristic-based optimization of the \FJ plan
(factorization).  This has the advantage of reusing an existing
cost-based optimizer, but the disadvantage is that an integrated
optimizer may be able to find better plans.  Third, our current
optimizer does not make use of existing indices.  It is known that the
optimization problem for join plans becomes harder in the presence of
foreign key indices~\cite{DBLP:journals/pvldb/LeisGMBK015}, and we
expect the same to hold for \FJ plans.  All these limitations require
future work.  As we have designed \FJ to closely capture binary join
while also generalizing it, we hope the solutions to these problems
can also be smoothly transferred from binary join to \FJ.

Finally, we have made several observations during this project, some
of them quite surprising (to us), which we believe deserve a future
study.  We observed that a major bottleneck is the materialization of
intermediate results in bushy plans; an improved materialization
algorithm may speed up \FJ on bushy plans.  One promising idea is to
be more aggressively lazy and keep \COLTs unexpanded during
materialization, which essentially leads to a factorized
representation of the intermediates.  We also observed that, contrary
to common belief, a cyclic query does not necessarily mean \WCOJ
algorithms are faster, and an acyclic query does not mean they are
slow.  A natural question is thus ``when exactly are \WCOJ algorithms
faster than binary join?''  Answering this question will also help us
design a better optimizer for \FJ.  The optimizer can output a plan
closer to \WCOJ when it expects major speedups.  We note that the
query optimizer by~\cite{DBLP:journals/pvldb/FreitagBSKN20} switches
between \GJ and binary join depending on the estimated cardinality.
In contrast, an optimizer for \FJ should smoothly transform a \FJ plan
to fully explore the design space between the two extremes of binary
join and \GJ.  Finally, we realized that, rather surprisingly, \GJ and
traditional joins diverge in their choice of the inner table (called
the cover in our paper): \GJ requires that to be the smallest
(otherwise it is not optimal), while a traditional plan will chose it
to be the largest (to save the cost of computing its hash table).
Future work is required for a better informed decision for the choice
of the inner relation.

\balance

\bibliographystyle{ACM-Reference-Format}
\bibliography{references}


\begin{thebibliography}{28}


\ifx \showCODEN    \undefined \def \showCODEN     #1{\unskip}     \fi
\ifx \showDOI      \undefined \def \showDOI       #1{#1}\fi
\ifx \showISBNx    \undefined \def \showISBNx     #1{\unskip}     \fi
\ifx \showISBNxiii \undefined \def \showISBNxiii  #1{\unskip}     \fi
\ifx \showISSN     \undefined \def \showISSN      #1{\unskip}     \fi
\ifx \showLCCN     \undefined \def \showLCCN      #1{\unskip}     \fi
\ifx \shownote     \undefined \def \shownote      #1{#1}          \fi
\ifx \showarticletitle \undefined \def \showarticletitle #1{#1}   \fi
\ifx \showURL      \undefined \def \showURL       {\relax}        \fi
\providecommand\bibfield[2]{#2}
\providecommand\bibinfo[2]{#2}
\providecommand\natexlab[1]{#1}
\providecommand\showeprint[2][]{arXiv:#2}

\bibitem[Abadi et~al\mbox{.}(2013)]%
        {DBLP:journals/ftdb/AbadiBHIM13}
\bibfield{author}{\bibinfo{person}{Daniel Abadi}, \bibinfo{person}{Peter~A.
  Boncz}, \bibinfo{person}{Stavros Harizopoulos}, \bibinfo{person}{Stratos
  Idreos}, {and} \bibinfo{person}{Samuel Madden}.}
  \bibinfo{year}{2013}\natexlab{}.
\newblock \showarticletitle{The Design and Implementation of Modern
  Column-Oriented Database Systems}.
\newblock \bibinfo{journal}{\emph{Found. Trends Databases}}
  \bibinfo{volume}{5}, \bibinfo{number}{3} (\bibinfo{year}{2013}),
  \bibinfo{pages}{197--280}.
\newblock
\urldef\tempurl%
\url{https://doi.org/10.1561/1900000024}
\showDOI{\tempurl}


\bibitem[Aberger et~al\mbox{.}(2017)]%
        {DBLP:journals/tods/AbergerLTNOR17}
\bibfield{author}{\bibinfo{person}{Christopher~R. Aberger},
  \bibinfo{person}{Andrew Lamb}, \bibinfo{person}{Susan Tu},
  \bibinfo{person}{Andres N{\"{o}}tzli}, \bibinfo{person}{Kunle Olukotun},
  {and} \bibinfo{person}{Christopher R{\'{e}}}.}
  \bibinfo{year}{2017}\natexlab{}.
\newblock \showarticletitle{EmptyHeaded: {A} Relational Engine for Graph
  Processing}.
\newblock \bibinfo{journal}{\emph{{ACM} Trans. Database Syst.}}
  \bibinfo{volume}{42}, \bibinfo{number}{4} (\bibinfo{year}{2017}),
  \bibinfo{pages}{20:1--20:44}.
\newblock
\urldef\tempurl%
\url{https://doi.org/10.1145/3129246}
\showDOI{\tempurl}


\bibitem[Atserias et~al\mbox{.}(2013)]%
        {DBLP:journals/siamcomp/AtseriasGM13}
\bibfield{author}{\bibinfo{person}{Albert Atserias}, \bibinfo{person}{Martin
  Grohe}, {and} \bibinfo{person}{D{\'{a}}niel Marx}.}
  \bibinfo{year}{2013}\natexlab{}.
\newblock \showarticletitle{Size Bounds and Query Plans for Relational Joins}.
\newblock \bibinfo{journal}{\emph{{SIAM} J. Comput.}} \bibinfo{volume}{42},
  \bibinfo{number}{4} (\bibinfo{year}{2013}), \bibinfo{pages}{1737--1767}.
\newblock
\urldef\tempurl%
\url{https://doi.org/10.1137/110859440}
\showDOI{\tempurl}


\bibitem[Avnur and Hellerstein(2000)]%
        {DBLP:conf/sigmod/HellersteinA00}
\bibfield{author}{\bibinfo{person}{Ron Avnur} {and} \bibinfo{person}{Joseph~M.
  Hellerstein}.} \bibinfo{year}{2000}\natexlab{}.
\newblock \showarticletitle{Eddies: Continuously Adaptive Query Processing}. In
  \bibinfo{booktitle}{\emph{Proceedings of the 2000 {ACM} {SIGMOD}
  International Conference on Management of Data, May 16-18, 2000, Dallas,
  Texas, {USA}}}, \bibfield{editor}{\bibinfo{person}{Weidong Chen},
  \bibinfo{person}{Jeffrey~F. Naughton}, {and} \bibinfo{person}{Philip~A.
  Bernstein}} (Eds.). \bibinfo{publisher}{{ACM}}, \bibinfo{pages}{261--272}.
\newblock
\urldef\tempurl%
\url{https://doi.org/10.1145/342009.335420}
\showDOI{\tempurl}


\bibitem[Fagin(1983)]%
        {DBLP:journals/jacm/Fagin83}
\bibfield{author}{\bibinfo{person}{Ronald Fagin}.}
  \bibinfo{year}{1983}\natexlab{}.
\newblock \showarticletitle{Degrees of Acyclicity for Hypergraphs and
  Relational Database Schemes}.
\newblock \bibinfo{journal}{\emph{J. {ACM}}} \bibinfo{volume}{30},
  \bibinfo{number}{3} (\bibinfo{year}{1983}), \bibinfo{pages}{514--550}.
\newblock
\urldef\tempurl%
\url{https://doi.org/10.1145/2402.322390}
\showDOI{\tempurl}


\bibitem[Feng et~al\mbox{.}(2023)]%
        {kuzu:cidr}
\bibfield{author}{\bibinfo{person}{Xiyang Feng}, \bibinfo{person}{Guodong Jin},
  \bibinfo{person}{Ziyi Chen}, \bibinfo{person}{Chang Liu}, {and}
  \bibinfo{person}{Semih Saliho\u{g}lu}.} \bibinfo{year}{2023}\natexlab{}.
\newblock \showarticletitle{K\`uzu Graph Database Management System}. In
  \bibinfo{booktitle}{\emph{CIDR}}.
\newblock


\bibitem[Freitag et~al\mbox{.}(2020)]%
        {DBLP:journals/pvldb/FreitagBSKN20}
\bibfield{author}{\bibinfo{person}{Michael~J. Freitag},
  \bibinfo{person}{Maximilian Bandle}, \bibinfo{person}{Tobias Schmidt},
  \bibinfo{person}{Alfons Kemper}, {and} \bibinfo{person}{Thomas Neumann}.}
  \bibinfo{year}{2020}\natexlab{}.
\newblock \showarticletitle{Adopting Worst-Case Optimal Joins in Relational
  Database Systems}.
\newblock \bibinfo{journal}{\emph{Proc. {VLDB} Endow.}} \bibinfo{volume}{13},
  \bibinfo{number}{11} (\bibinfo{year}{2020}), \bibinfo{pages}{1891--1904}.
\newblock
\urldef\tempurl%
\url{http://www.vldb.org/pvldb/vol13/p1891-freitag.pdf}
\showURL{%
\tempurl}


\bibitem[Graefe(1993)]%
        {DBLP:journals/csur/Graefe93}
\bibfield{author}{\bibinfo{person}{Goetz Graefe}.}
  \bibinfo{year}{1993}\natexlab{}.
\newblock \showarticletitle{Query Evaluation Techniques for Large Databases}.
\newblock \bibinfo{journal}{\emph{{ACM} Comput. Surv.}} \bibinfo{volume}{25},
  \bibinfo{number}{2} (\bibinfo{year}{1993}), \bibinfo{pages}{73--170}.
\newblock
\urldef\tempurl%
\url{https://doi.org/10.1145/152610.152611}
\showDOI{\tempurl}


\bibitem[Graefe et~al\mbox{.}(1998)]%
        {DBLP:conf/vldb/GraefeBC98}
\bibfield{author}{\bibinfo{person}{Goetz Graefe}, \bibinfo{person}{Ross
  Bunker}, {and} \bibinfo{person}{Shaun Cooper}.}
  \bibinfo{year}{1998}\natexlab{}.
\newblock \showarticletitle{Hash Joins and Hash Teams in Microsoft {SQL}
  Server}. In \bibinfo{booktitle}{\emph{VLDB'98, Proceedings of 24rd
  International Conference on Very Large Data Bases, August 24-27, 1998, New
  York City, New York, {USA}}}, \bibfield{editor}{\bibinfo{person}{Ashish
  Gupta}, \bibinfo{person}{Oded Shmueli}, {and} \bibinfo{person}{Jennifer
  Widom}} (Eds.). \bibinfo{publisher}{Morgan Kaufmann},
  \bibinfo{pages}{86--97}.
\newblock
\urldef\tempurl%
\url{http://www.vldb.org/conf/1998/p086.pdf}
\showURL{%
\tempurl}


\bibitem[Idreos et~al\mbox{.}(2007a)]%
        {DBLP:conf/cidr/IdreosKM07}
\bibfield{author}{\bibinfo{person}{Stratos Idreos}, \bibinfo{person}{Martin~L.
  Kersten}, {and} \bibinfo{person}{Stefan Manegold}.}
  \bibinfo{year}{2007}\natexlab{a}.
\newblock \showarticletitle{Database Cracking}. In
  \bibinfo{booktitle}{\emph{Third Biennial Conference on Innovative Data
  Systems Research, {CIDR} 2007, Asilomar, CA, USA, January 7-10, 2007, Online
  Proceedings}}. \bibinfo{publisher}{www.cidrdb.org}, \bibinfo{pages}{68--78}.
\newblock
\urldef\tempurl%
\url{http://cidrdb.org/cidr2007/papers/cidr07p07.pdf}
\showURL{%
\tempurl}


\bibitem[Idreos et~al\mbox{.}(2007b)]%
        {DBLP:conf/sigmod/IdreosKM07}
\bibfield{author}{\bibinfo{person}{Stratos Idreos}, \bibinfo{person}{Martin~L.
  Kersten}, {and} \bibinfo{person}{Stefan Manegold}.}
  \bibinfo{year}{2007}\natexlab{b}.
\newblock \showarticletitle{Updating a cracked database}. In
  \bibinfo{booktitle}{\emph{Proceedings of the {ACM} {SIGMOD} International
  Conference on Management of Data, Beijing, China, June 12-14, 2007}},
  \bibfield{editor}{\bibinfo{person}{Chee~Yong Chan},
  \bibinfo{person}{Beng~Chin Ooi}, {and} \bibinfo{person}{Aoying Zhou}} (Eds.).
  \bibinfo{publisher}{{ACM}}, \bibinfo{pages}{413--424}.
\newblock
\urldef\tempurl%
\url{https://doi.org/10.1145/1247480.1247527}
\showDOI{\tempurl}


\bibitem[Kemper et~al\mbox{.}(1999)]%
        {DBLP:conf/vldb/KemperKW99}
\bibfield{author}{\bibinfo{person}{Alfons Kemper}, \bibinfo{person}{Donald
  Kossmann}, {and} \bibinfo{person}{Christian Wiesner}.}
  \bibinfo{year}{1999}\natexlab{}.
\newblock \showarticletitle{Generalised Hash Teams for Join and Group-by}. In
  \bibinfo{booktitle}{\emph{VLDB'99, Proceedings of 25th International
  Conference on Very Large Data Bases, September 7-10, 1999, Edinburgh,
  Scotland, {UK}}}, \bibfield{editor}{\bibinfo{person}{Malcolm~P. Atkinson},
  \bibinfo{person}{Maria~E. Orlowska}, \bibinfo{person}{Patrick Valduriez},
  \bibinfo{person}{Stanley~B. Zdonik}, {and} \bibinfo{person}{Michael~L.
  Brodie}} (Eds.). \bibinfo{publisher}{Morgan Kaufmann},
  \bibinfo{pages}{30--41}.
\newblock
\urldef\tempurl%
\url{http://www.vldb.org/conf/1999/P3.pdf}
\showURL{%
\tempurl}


\bibitem[Kersten et~al\mbox{.}(2018)]%
        {DBLP:journals/pvldb/KerstenLKNPB18}
\bibfield{author}{\bibinfo{person}{Timo Kersten}, \bibinfo{person}{Viktor
  Leis}, \bibinfo{person}{Alfons Kemper}, \bibinfo{person}{Thomas Neumann},
  \bibinfo{person}{Andrew Pavlo}, {and} \bibinfo{person}{Peter~A. Boncz}.}
  \bibinfo{year}{2018}\natexlab{}.
\newblock \showarticletitle{Everything You Always Wanted to Know About Compiled
  and Vectorized Queries But Were Afraid to Ask}.
\newblock \bibinfo{journal}{\emph{Proc. {VLDB} Endow.}} \bibinfo{volume}{11},
  \bibinfo{number}{13} (\bibinfo{year}{2018}), \bibinfo{pages}{2209--2222}.
\newblock
\urldef\tempurl%
\url{https://doi.org/10.14778/3275366.3275370}
\showDOI{\tempurl}


\bibitem[Khamis et~al\mbox{.}(2017)]%
        {DBLP:conf/pods/Khamis0S17}
\bibfield{author}{\bibinfo{person}{Mahmoud~Abo Khamis},
  \bibinfo{person}{Hung~Q. Ngo}, {and} \bibinfo{person}{Dan Suciu}.}
  \bibinfo{year}{2017}\natexlab{}.
\newblock \showarticletitle{What Do Shannon-type Inequalities, Submodular
  Width, and Disjunctive Datalog Have to Do with One Another?}. In
  \bibinfo{booktitle}{\emph{Proceedings of the 36th {ACM} {SIGMOD-SIGACT-SIGAI}
  Symposium on Principles of Database Systems, {PODS} 2017, Chicago, IL, USA,
  May 14-19, 2017}}, \bibfield{editor}{\bibinfo{person}{Emanuel Sallinger},
  \bibinfo{person}{Jan~Van den Bussche}, {and} \bibinfo{person}{Floris Geerts}}
  (Eds.). \bibinfo{publisher}{{ACM}}, \bibinfo{pages}{429--444}.
\newblock
\urldef\tempurl%
\url{https://doi.org/10.1145/3034786.3056105}
\showDOI{\tempurl}


\bibitem[Leis et~al\mbox{.}(2015)]%
        {DBLP:journals/pvldb/LeisGMBK015}
\bibfield{author}{\bibinfo{person}{Viktor Leis}, \bibinfo{person}{Andrey
  Gubichev}, \bibinfo{person}{Atanas Mirchev}, \bibinfo{person}{Peter~A.
  Boncz}, \bibinfo{person}{Alfons Kemper}, {and} \bibinfo{person}{Thomas
  Neumann}.} \bibinfo{year}{2015}\natexlab{}.
\newblock \showarticletitle{How Good Are Query Optimizers, Really?}
\newblock \bibinfo{journal}{\emph{Proc. {VLDB} Endow.}} \bibinfo{volume}{9},
  \bibinfo{number}{3} (\bibinfo{year}{2015}), \bibinfo{pages}{204--215}.
\newblock
\urldef\tempurl%
\url{https://doi.org/10.14778/2850583.2850594}
\showDOI{\tempurl}


\bibitem[Mhedhbi et~al\mbox{.}(2021)]%
        {DBLP:conf/sigmod/MhedhbiLKWS21}
\bibfield{author}{\bibinfo{person}{Amine Mhedhbi}, \bibinfo{person}{Matteo
  Lissandrini}, \bibinfo{person}{Laurens Kuiper}, \bibinfo{person}{Jack
  Waudby}, {and} \bibinfo{person}{G{\'{a}}bor Sz{\'{a}}rnyas}.}
  \bibinfo{year}{2021}\natexlab{}.
\newblock \showarticletitle{{LSQB:} a large-scale subgraph query benchmark}. In
  \bibinfo{booktitle}{\emph{{GRADES-NDA} '21: Proceedings of the 4th {ACM}
  {SIGMOD} Joint International Workshop on Graph Data Management Experiences
  {\&} Systems {(GRADES)} and Network Data Analytics (NDA), Virtual Event,
  China, 20 June 2021}}, \bibfield{editor}{\bibinfo{person}{Vasiliki Kalavri}
  {and} \bibinfo{person}{Nikolay Yakovets}} (Eds.). \bibinfo{publisher}{{ACM}},
  \bibinfo{pages}{8:1--8:11}.
\newblock
\urldef\tempurl%
\url{https://doi.org/10.1145/3461837.3464516}
\showDOI{\tempurl}


\bibitem[Mhedhbi and Salihoglu(2019)]%
        {DBLP:journals/pvldb/MhedhbiS19}
\bibfield{author}{\bibinfo{person}{Amine Mhedhbi} {and} \bibinfo{person}{Semih
  Salihoglu}.} \bibinfo{year}{2019}\natexlab{}.
\newblock \showarticletitle{Optimizing Subgraph Queries by Combining Binary and
  Worst-Case Optimal Joins}.
\newblock \bibinfo{journal}{\emph{Proc. {VLDB} Endow.}} \bibinfo{volume}{12},
  \bibinfo{number}{11} (\bibinfo{year}{2019}), \bibinfo{pages}{1692--1704}.
\newblock
\urldef\tempurl%
\url{https://doi.org/10.14778/3342263.3342643}
\showDOI{\tempurl}


\bibitem[Neumann(2011)]%
        {DBLP:journals/pvldb/Neumann11}
\bibfield{author}{\bibinfo{person}{Thomas Neumann}.}
  \bibinfo{year}{2011}\natexlab{}.
\newblock \showarticletitle{Efficiently Compiling Efficient Query Plans for
  Modern Hardware}.
\newblock \bibinfo{journal}{\emph{Proc. {VLDB} Endow.}} \bibinfo{volume}{4},
  \bibinfo{number}{9} (\bibinfo{year}{2011}), \bibinfo{pages}{539--550}.
\newblock
\urldef\tempurl%
\url{https://doi.org/10.14778/2002938.2002940}
\showDOI{\tempurl}


\bibitem[Ngo(2018)]%
        {DBLP:conf/pods/000118}
\bibfield{author}{\bibinfo{person}{Hung~Q. Ngo}.}
  \bibinfo{year}{2018}\natexlab{}.
\newblock \showarticletitle{Worst-Case Optimal Join Algorithms: Techniques,
  Results, and Open Problems}. In \bibinfo{booktitle}{\emph{Proceedings of the
  37th {ACM} {SIGMOD-SIGACT-SIGAI} Symposium on Principles of Database Systems,
  Houston, TX, USA, June 10-15, 2018}},
  \bibfield{editor}{\bibinfo{person}{Jan~Van den Bussche} {and}
  \bibinfo{person}{Marcelo Arenas}} (Eds.). \bibinfo{publisher}{{ACM}},
  \bibinfo{address}{New York, NY, USA}, \bibinfo{pages}{111--124}.
\newblock
\urldef\tempurl%
\url{https://doi.org/10.1145/3196959.3196990}
\showDOI{\tempurl}


\bibitem[Ngo et~al\mbox{.}(2012)]%
        {DBLP:conf/pods/NgoPRR12}
\bibfield{author}{\bibinfo{person}{Hung~Q. Ngo}, \bibinfo{person}{Ely Porat},
  \bibinfo{person}{Christopher R{\'{e}}}, {and} \bibinfo{person}{Atri Rudra}.}
  \bibinfo{year}{2012}\natexlab{}.
\newblock \showarticletitle{Worst-case optimal join algorithms: [extended
  abstract]}. In \bibinfo{booktitle}{\emph{Proceedings of the 31st {ACM}
  {SIGMOD-SIGACT-SIGART} Symposium on Principles of Database Systems, {PODS}
  2012, Scottsdale, AZ, USA, May 20-24, 2012}},
  \bibfield{editor}{\bibinfo{person}{Michael Benedikt}, \bibinfo{person}{Markus
  Kr{\"{o}}tzsch}, {and} \bibinfo{person}{Maurizio Lenzerini}} (Eds.).
  \bibinfo{publisher}{{ACM}}, \bibinfo{address}{New York, NY, USA},
  \bibinfo{pages}{37--48}.
\newblock
\urldef\tempurl%
\url{https://doi.org/10.1145/2213556.2213565}
\showDOI{\tempurl}


\bibitem[Ngo et~al\mbox{.}(2013)]%
        {DBLP:journals/sigmod/NgoRR13}
\bibfield{author}{\bibinfo{person}{Hung~Q. Ngo}, \bibinfo{person}{Christopher
  R{\'{e}}}, {and} \bibinfo{person}{Atri Rudra}.}
  \bibinfo{year}{2013}\natexlab{}.
\newblock \showarticletitle{Skew strikes back: new developments in the theory
  of join algorithms}.
\newblock \bibinfo{journal}{\emph{{SIGMOD} Rec.}} \bibinfo{volume}{42},
  \bibinfo{number}{4} (\bibinfo{year}{2013}), \bibinfo{pages}{5--16}.
\newblock
\urldef\tempurl%
\url{https://doi.org/10.1145/2590989.2590991}
\showDOI{\tempurl}


\bibitem[Olteanu and Schleich(2016)]%
        {DBLP:journals/sigmod/OlteanuS16}
\bibfield{author}{\bibinfo{person}{Dan Olteanu} {and}
  \bibinfo{person}{Maximilian Schleich}.} \bibinfo{year}{2016}\natexlab{}.
\newblock \showarticletitle{Factorized Databases}.
\newblock \bibinfo{journal}{\emph{{SIGMOD} Rec.}} \bibinfo{volume}{45},
  \bibinfo{number}{2} (\bibinfo{year}{2016}), \bibinfo{pages}{5--16}.
\newblock
\urldef\tempurl%
\url{https://doi.org/10.1145/3003665.3003667}
\showDOI{\tempurl}


\bibitem[Padmanabhan et~al\mbox{.}(2001)]%
        {DBLP:conf/icde/PadmanabhanAMJ01}
\bibfield{author}{\bibinfo{person}{Sriram Padmanabhan},
  \bibinfo{person}{Timothy Malkemus}, \bibinfo{person}{Ramesh~C. Agarwal},
  {and} \bibinfo{person}{Anant Jhingran}.} \bibinfo{year}{2001}\natexlab{}.
\newblock \showarticletitle{Block Oriented Processing of Relational Database
  Operations in Modern Computer Architectures}. In
  \bibinfo{booktitle}{\emph{Proceedings of the 17th International Conference on
  Data Engineering, April 2-6, 2001, Heidelberg, Germany}},
  \bibfield{editor}{\bibinfo{person}{Dimitrios Georgakopoulos} {and}
  \bibinfo{person}{Alexander Buchmann}} (Eds.). \bibinfo{publisher}{{IEEE}
  Computer Society}, \bibinfo{pages}{567--574}.
\newblock
\urldef\tempurl%
\url{https://doi.org/10.1109/ICDE.2001.914871}
\showDOI{\tempurl}


\bibitem[Raasveldt(2022)]%
        {DBLP:conf/vldb/Raasveldt22}
\bibfield{author}{\bibinfo{person}{Mark Raasveldt}.}
  \bibinfo{year}{2022}\natexlab{}.
\newblock \showarticletitle{DuckDB - {A} Modern Modular and Extensible Database
  System}. In \bibinfo{booktitle}{\emph{1st International Workshop on
  Composable Data Management Systems, CDMS@VLDB 2022, Sydney, Australia,
  September 9, 2022}}, \bibfield{editor}{\bibinfo{person}{Satyanarayana~R.
  Valluri} {and} \bibinfo{person}{Mohamed Zait}} (Eds.).
\newblock
\urldef\tempurl%
\url{https://cdmsworkshop.github.io/2022/Proceedings/Keynotes/Abstract\_MarkRaasveldt.pdf}
\showURL{%
\tempurl}


\bibitem[Raasveldt and M{\"{u}}hleisen(2020)]%
        {DBLP:conf/cidr/RaasveldtM20}
\bibfield{author}{\bibinfo{person}{Mark Raasveldt} {and}
  \bibinfo{person}{Hannes M{\"{u}}hleisen}.} \bibinfo{year}{2020}\natexlab{}.
\newblock \showarticletitle{Data Management for Data Science - Towards Embedded
  Analytics}. In \bibinfo{booktitle}{\emph{10th Conference on Innovative Data
  Systems Research, {CIDR} 2020, Amsterdam, The Netherlands, January 12-15,
  2020, Online Proceedings}}. \bibinfo{publisher}{www.cidrdb.org}.
\newblock
\urldef\tempurl%
\url{http://cidrdb.org/cidr2020/papers/p23-raasveldt-cidr20.pdf}
\showURL{%
\tempurl}


\bibitem[Selinger et~al\mbox{.}(1979)]%
        {DBLP:conf/sigmod/SelingerACLP79}
\bibfield{author}{\bibinfo{person}{Patricia~G. Selinger},
  \bibinfo{person}{Morton~M. Astrahan}, \bibinfo{person}{Donald~D. Chamberlin},
  \bibinfo{person}{Raymond~A. Lorie}, {and} \bibinfo{person}{Thomas~G. Price}.}
  \bibinfo{year}{1979}\natexlab{}.
\newblock \showarticletitle{Access Path Selection in a Relational Database
  Management System}. In \bibinfo{booktitle}{\emph{Proceedings of the 1979
  {ACM} {SIGMOD} International Conference on Management of Data, Boston,
  Massachusetts, USA, May 30 - June 1}},
  \bibfield{editor}{\bibinfo{person}{Philip~A. Bernstein}} (Ed.).
  \bibinfo{publisher}{{ACM}}, \bibinfo{pages}{23--34}.
\newblock
\urldef\tempurl%
\url{https://doi.org/10.1145/582095.582099}
\showDOI{\tempurl}


\bibitem[Veldhuizen(2014)]%
        {DBLP:conf/icdt/Veldhuizen14}
\bibfield{author}{\bibinfo{person}{Todd~L. Veldhuizen}.}
  \bibinfo{year}{2014}\natexlab{}.
\newblock \showarticletitle{Triejoin: {A} Simple, Worst-Case Optimal Join
  Algorithm}. In \bibinfo{booktitle}{\emph{Proc. 17th International Conference
  on Database Theory (ICDT), Athens, Greece, March 24-28, 2014}},
  \bibfield{editor}{\bibinfo{person}{Nicole Schweikardt},
  \bibinfo{person}{Vassilis Christophides}, {and} \bibinfo{person}{Vincent
  Leroy}} (Eds.). \bibinfo{publisher}{OpenProceedings.org},
  \bibinfo{address}{Athens, Greece}, \bibinfo{pages}{96--106}.
\newblock
\urldef\tempurl%
\url{https://doi.org/10.5441/002/icdt.2014.13}
\showDOI{\tempurl}


\bibitem[Yannakakis(1981)]%
        {DBLP:conf/vldb/Yannakakis81}
\bibfield{author}{\bibinfo{person}{Mihalis Yannakakis}.}
  \bibinfo{year}{1981}\natexlab{}.
\newblock \showarticletitle{Algorithms for Acyclic Database Schemes}. In
  \bibinfo{booktitle}{\emph{Very Large Data Bases, 7th International
  Conference, September 9-11, 1981, Cannes, France, Proceedings}}.
  \bibinfo{publisher}{{IEEE} Computer Society}, \bibinfo{pages}{82--94}.
\newblock


\end{thebibliography}



\end{document}